\documentclass[aps,a4paper,twocolumn,floatfix,showkeys,
groupaddress]{revtex4}

\usepackage{bm,amsmath,amssymb,amsfonts}

\usepackage[dvips]{graphicx}
\usepackage[colorlinks=true,citecolor=magenta,urlcolor=blue,filecolor=red]{hyperref}
\usepackage{color}
\definecolor{mycolor}{rgb}{0.7,0.3,0}
\begin{document}
	
	\title{\textcolor{mycolor}
		{Localization, transport, flux induced extended modes and mobility edge in a self-similar corral substrate }}  
	
\author{Sayan Bhattacharya}
\email{sayanbhatta2002@gmail.com}
\affiliation{Department of Physics, Acharya Prafulla Chandra College,
	New Barrackpore, Kolkata, West Bengal-700 131, India}

\author{Rhiddha Acharjee}
\email{rhiddhaacharjee777@gmail.com}
\affiliation{Department of Physics, University of Calcutta, 92,
	Acharya Prafulla Chandra Road, Kolkata, West Bengal-700 009, India}

\author{Atanu Nandy}
\email{atanunandy1989@gmail.com}
\affiliation{Department of Physics, Acharya Prafulla Chandra College,
	New Barrackpore, Kolkata, West Bengal-700 131, India}

	\begin{abstract}
		We address that a single-band tight-binding Hamiltonian defined on a self-similar corral substrate can give rise to a set of non-diffusive localized modes that follow the same hierarchical distribution. As the lattice, the spatial extent of quantum prison containing a cluster of atomic sites is dependent on the generation of fractal structure. Apart from the quantum imprisonment of the excitation, a magnetic flux threading each elementary plaquette is shown to destroy the boundedness and generate an absolutely continuous sub-band populated by resonant eigen functions. Flux induced engineering of quantum states is corroborated through the evaluation of inverse participation ratio and quantum transport. Moreover, the robustness of the extended states has been checked in presence of diagonal disorder and off-diagonal anisotropy. Flux modulated single-particle mobility edge is characterized through mutlifractal analysis. Quantum interference is the essential issue, reported here, that manipulates the kinematics of the excitation and this is manifested by the workout of persistent current. 
	\end{abstract}
	\maketitle


\begin{figure*}[ht]
	\centering
	
	(a){\includegraphics[width=0.28\textwidth]{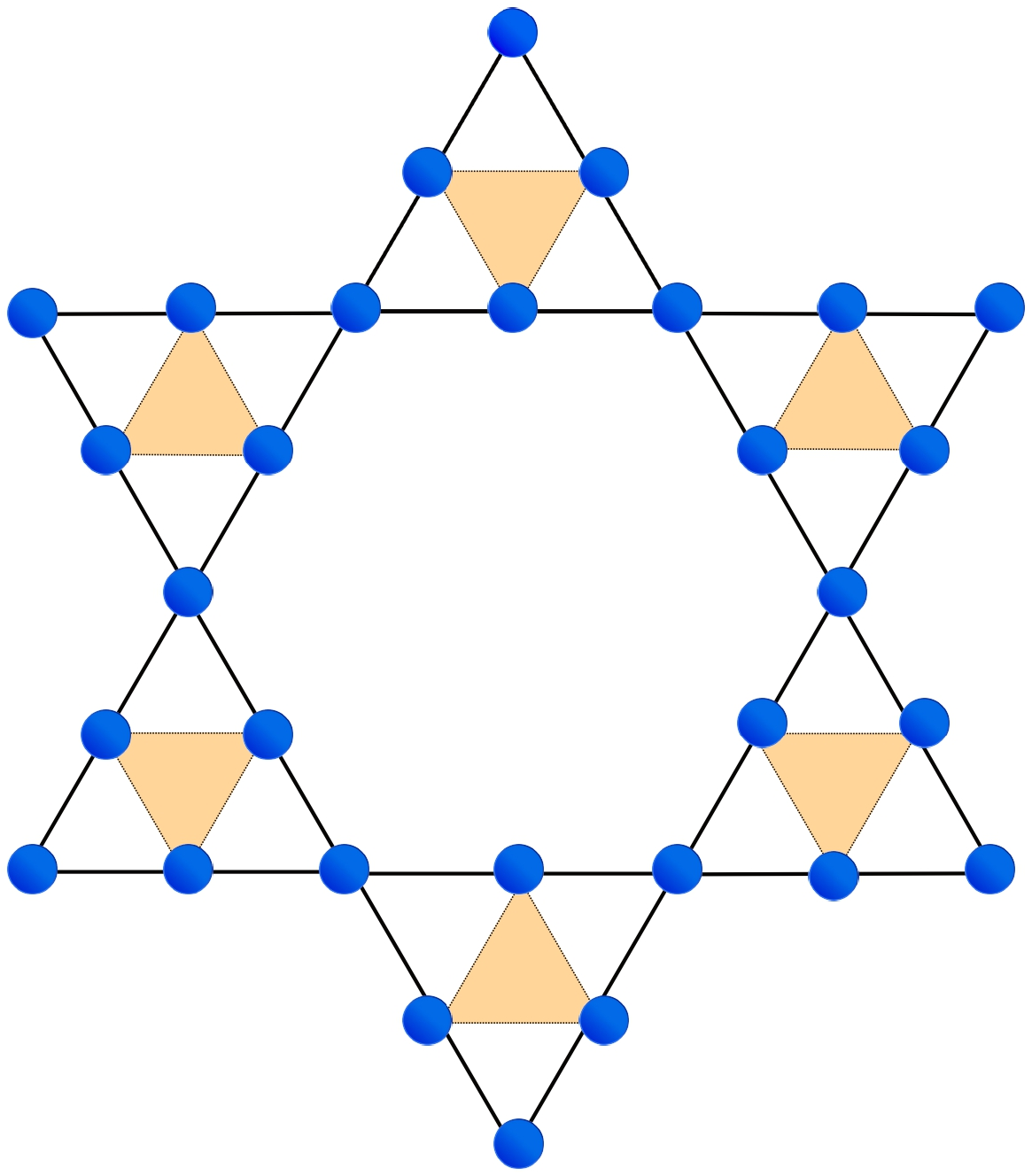}}
	\hspace{0.15cm}
	(b){\includegraphics[width=0.28\textwidth]{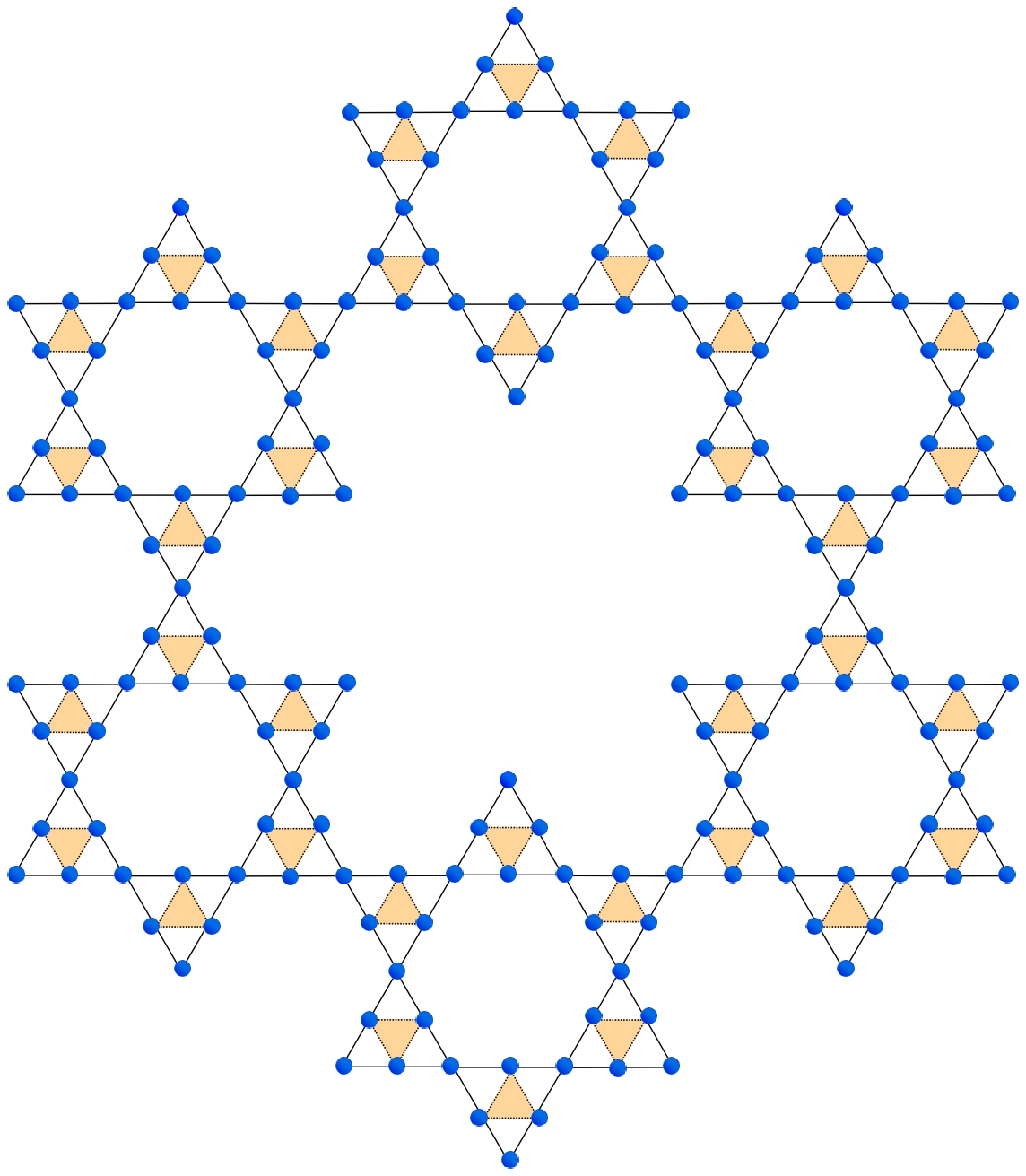}}
	\hspace{0.15cm}
	(c){\includegraphics[width=0.30\textwidth]{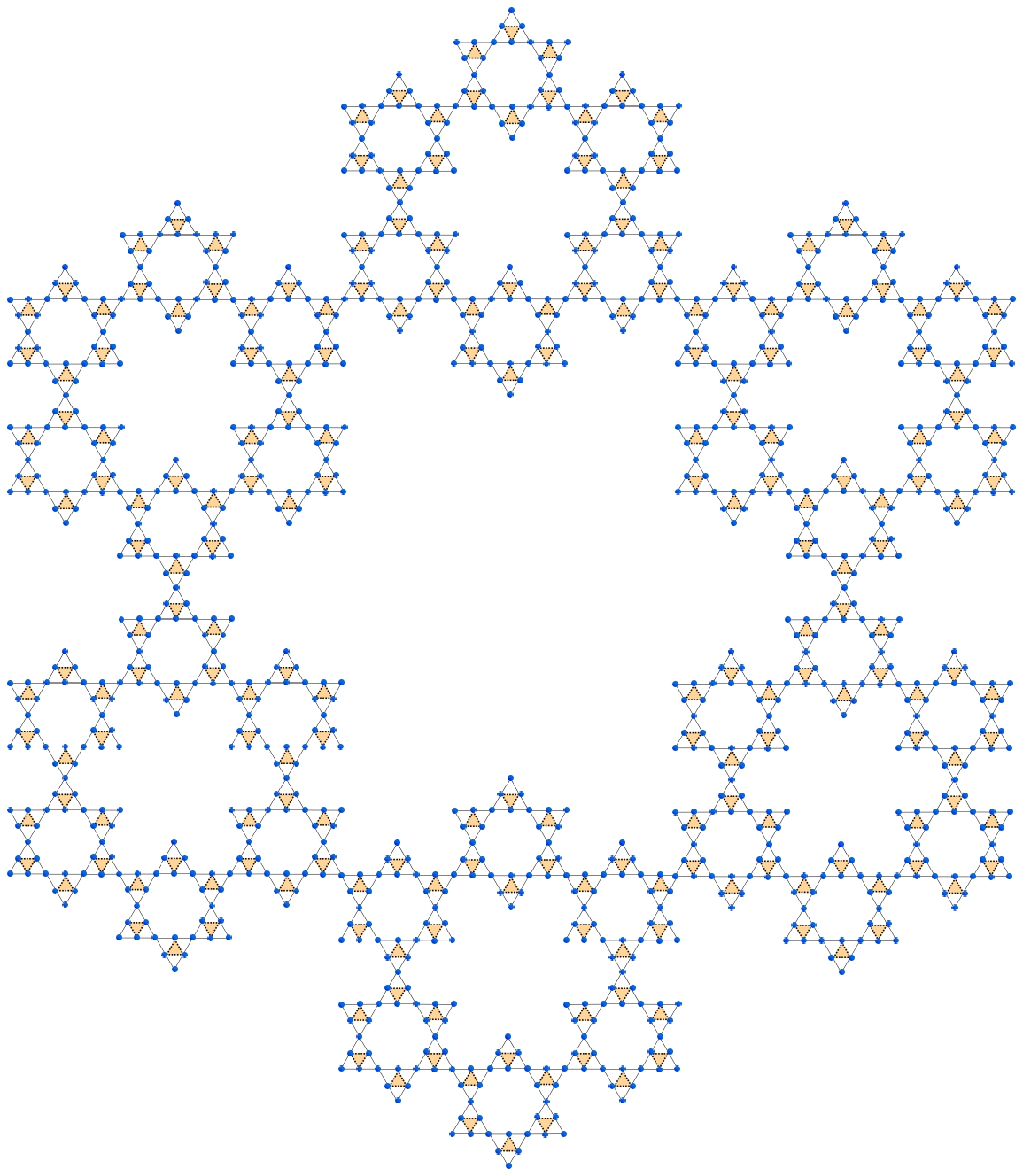}}
	
	\caption{Germination scheme of the self-similar hexaflake corral from a basic motif. The pictorial representation of (a) first generation, (b) second generation, and (c) third generation corral respectively.}
   \label{corral}
\end{figure*}

	
\section{Introduction}
\label{intro}
The occurrence of localization of excitation in disordered lattices has received numerous recognition in the condensed matter community since it was first projected by P. W. Anderson~\cite{ander} 
for uncorrelated random arrangement of potentials.
The `spatial extension' of an eigenstate falls off in an exponential fashion
with a well-defined notion of a characteristic localization length.
The resulting localization effect in presence of disorder is highly dependent on the dimensionality of the lattice, the kind of lattice topology and also on the typical nature of the potential taken under consideration.
Analysis of different tight-binding lattice models in three, two, and 
one dimensions unfold that, while there is a critical concentration of the disorder in three dimensions to observe the absence of diffusive states~\cite{ander}, in two dimensions~\cite{ander2} and in one dimension~\cite{bor,dey} all single-particle modes are localized in general, irrespective of the strength of the disorder. 
The interest in the groundbreaking Anderson localization (AL) has gained
considerable momentum because of
its experimental realization in ultracold atomic gases~\cite{billy}.
The particular topic carries everlasting momentum as the manifestation of disorder is ubiquitous~\cite{kramer,van} and 
is experimentally visualized in recent past for a wide variety of systems with 
the help of light~\cite{mp}-\cite{segev}, Bose-Einstein condensates~\cite{roati}, 
or the quite recently one involving cold atoms~\cite{white}, to name a few.

Deterministic fractal geometries are neither perfectly ordered systems nor completely disordered structures but they carry somewhat intermediate spectral flavor. Unlike the canonical case of AL, in such deterministic fractals, the localization occurs due to the
finite ramifications and self-similar fashion.
They are different in respect of their dimensionality from the homogeneous euclidean objects. Such networks are interesting playgrounds to explore the spectral properties because of non-integer dimensionality.
In fact, single-particle AL was tested in fractal and bifractal setups~\cite{schri}-\cite{sacha}.
Over the past few years fractal lattices have set up a benchmark because of their footprint on several physical directions like topological feature~\cite{cook,prem}
of single-particle eigenstate on fractals, momentum independent non-dispersive bands~\cite{bp1}-\cite{sb}
in several scale invariant fractal lattices
and Hall conductivity~\cite{hoo,yuan}, to name a few.
Electrons and phonons travelling through this type of non-periodic lattices undergo the non-integer dimension of
the geometries, making impression of the self-similar patterns on the corresponding
band structures, density of states (DOS), and the response.
Molecular chains based on carbon Sierpinski
triangle architectures~\cite{lage} have been addressed recently that cites the representation of
different spatial charge distribution which can be highlighted in
functional nano-devices.
The progress of experimental procedures
in synthesizing finite fractal molecules, like self-assembly~\cite{shang} has added additional momentum in this realm. 
Moreover, the interesting development in the fabrication mechanism has made possible to explore several challenging and intriguing physics on the fractal substrate. Recent literature has inspiring evidences of designing quantum fractal with the aid of atomic manipulation in a
scanning tunnelling microscope~\cite{slot}, fractal-based photonic lattice comprising of waveguides experimentally idealized with the help of standard femtosecond-
laser-writing technology~\cite{lust}.

Motivated by the different experimental highlights, recent years have witnessed a number of theoretical findings that have drawn remarkable attention to the condensed matter physics community. This includes researches on well-known level statistics~\cite{yori} in fractal clusters,
engineering quantum states~\cite{bp2,bp3}, 
Josephson effect provided by a fractal geometry~\cite{amun}, and stability of loop current states in fractal objects built with Bose-Einstein
condensates~\cite{koch}. etc.
We hope that this advancements in the journey of experimental strategy will give support to the different theoretical observations in future.

In this communication, we address a snowflake-like
quantum corral having a hexagonal symmetry~\cite{science}.
In the tight-binding workout, this may be considered as a hexagonal triangular
Sierpinski corral. Such quasi-one dimensional fractal geometries can be synthesized as reported by L. L. Lage et. al.~\cite{carbon} and quantum corrals have now become perfect candidates for studying electronic properties and exotic quantum response.
Our preliminary point of interest is to study the spectral response offered by this \textit{self-similar} quantum corral with the energy of the injected projectile. Absence of periodicity incidentally reveals a number of localized states separated by gaps constituting the electronic spectrum. However, creation of diffusive eigenstates by increasing the accessibility in such fractal lattices~\cite{acsimplex}, especially an absolutely
continuous band of the same, is quite non-trivial. Different correlation between the parameters can generate perfectly resonant Bloch-band leading to a complete escape from localization as reported~\cite{acnew} from our group.
But, here a uniform magnetic flux is shown to bring this prominent spectral modification in this \textit{quantum corral} structure, which is practically an addressed feature, we believe. 
The evolution of diffusive band is found to be very much stable even in the application of random diagonal disorder. The transmittivity also supports this robustness of the modes. This stability may direct one to test flux induced engineering of quantum states.
It is needless to say that flux here plays a crucial role in changing the dynamics of the electron. As a second motivation, interesting spectral competition between the external flux induced resonance and the structural aperiodicity makes us curious to search for single-particle mobility edge, if any. For this purpose, we have performed a detailed analysis including the quantum transport and inverse participation ratio. All these workouts provide a clear signal of existences of two different quantum phases in the spectrum. 
In fact, multifractal analysis (MFA) at this point
readily validates the possibility of single-particle mobility edge and makes our claim true.
Our analysis ends with the pictorial demonstration of quantum interference effect through the evaluation of persistent current. 

We thus summarize our findings at a glance.In Sec.~\ref{model} we first demonstrate the quantum corral using the tight-binding formalism. 
This section also highlights the presence of localized modes in the absence of an external flux. Sec.~\ref{mag} discusses the flux-dependent allowed eigenspectrum along with the evaluation of the 
density of states and the inverse participation ratio. Sec.~\ref{flux} covers the flux-tunable 
modifications of the response, where the flux-induced delocalization is corroborated through 
the study of quantum transport. Sec.~\ref{rob} examines the sustainability of the resonant band 
populated by diffusive eigenfunctions under the application of diagonal disorder or 
off-diagonal anisotropy. Sec.~\ref{mfan} introduces the appearance of a single-particle mobility 
edge in the electronic spectrum in the presence of flux, which is validated by a multifractal 
analysis. In Sec.~\ref{persi} we report the variation of the persistent current with flux and its 
tunability through internal parametric modulation. Sec.~\ref{phon} discusses the possible extension to photonics. Finally, in Sec.~\ref{closing} we draw our conclusions.

	\section{Model system and Hamiltonian}
	\label{model}
	
	\subsection{Description of the Hamiltonian}
We begin with the theoretical model of a self-similar corral structure~\cite{carbon} that follows a certain growth pattern. The systematic germination of the network is cited in Figure~\ref{corral}.As mentioned earlier, the rapid development of synthesis mechanism, this kind of aperiodic lattice can be modeled to study the nature of the single particle state and the response of the system. In our work, the underlying corral can be generated by translating Sierpiński triangles connected to form the self similar object. The single particle eigenstates in such quantum corral lattice can be mathematically manifested by using the following tight binding Hamiltonian, viz., 

\begin{equation}
	\mathcal{H} = \sum_{j} \epsilon_j c_j^\dagger c_j + \sum_{\langle j,k \rangle} \left( t_{jk} e^{i \theta_{jk}} c_j^\dagger c_k + \text{h.c.} \right),
\end{equation}

where the first term that precisely speaks for the potential information, is termed as the on-site energy in the tight binding description. The second term that carries the kinetic signature is known as the overlap integral between the nearest neighbouring atomic sites in the corral geometry. $c_j^\dagger (c_j)$ are the standard creation (annihilation) operators. The exponential factor associated with the off-diagonal term $t_{jk}$ of the Hamiltonian signifies the usual Peierls phase. This is the inheritable consequence of the Aharonov--Bohm (AB) effect and comes into play when an electron follows a closed trajectory that traps a finite magnetic flux $\Phi$ (in units of $\Phi_0 = \frac{hc}{e}$). The wave function of the electron acquires a phase and this has a sensitive impact on the overall spectral features offered by the system. The bulk sites have the coordination number equal to four, whereas the corner sites are associated with 2 nearest neighbouring sites. Depending upon the coordination number, we may put different on-site potential but their numerical values can be uniformly set as zero for all the sites, without any loss of generality and that of hopping can be put as unity throughout our analysis. Since we are interested in the lattice topology induced spectral modification, such choice of the parameters will not hamper the entire working procedure, physics-wise, to obtain relevant information. The difference equation (discretized form of the Schrodinger's equation) for the corral with tight binding form reads as, 
\begin{equation}
	 (E - \epsilon_j)\psi_j = \sum_{k}t_{jk}\psi_k
	 \label{dif}
\end{equation}

With this initial demonstration of the Hamiltonian, written in the (Wannier) basis, we now try to achieve some fundamental insight into the spectrum of this self similar quantum corral even in presence of definite magnetic flux $\Phi$. The magnetic flux is incorporated in each elementary Sierpiński triangle and hence the Aharonov--Bohm phase is $\theta = \frac{2\pi\Phi}{3\Phi_0}$. The circulation direction of the magnetic vector potential is taken in the clockwise direction.

		\subsection{Localized states in absence of flux}
	
		\begin{figure}[ht]
		
		(a)\includegraphics[clip,width=0.30\textwidth]{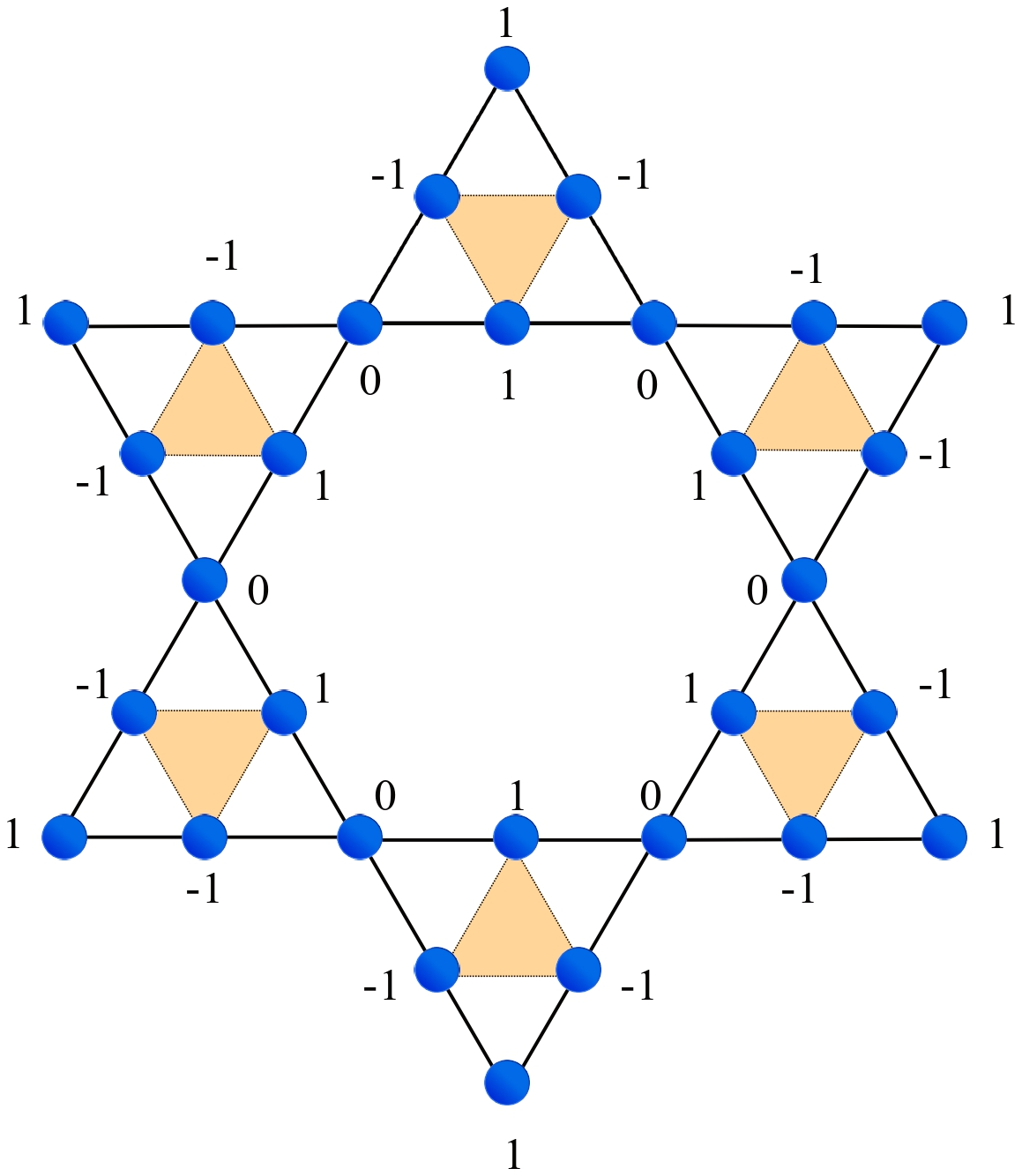}
		(b)\includegraphics[clip,width=0.38\textwidth]{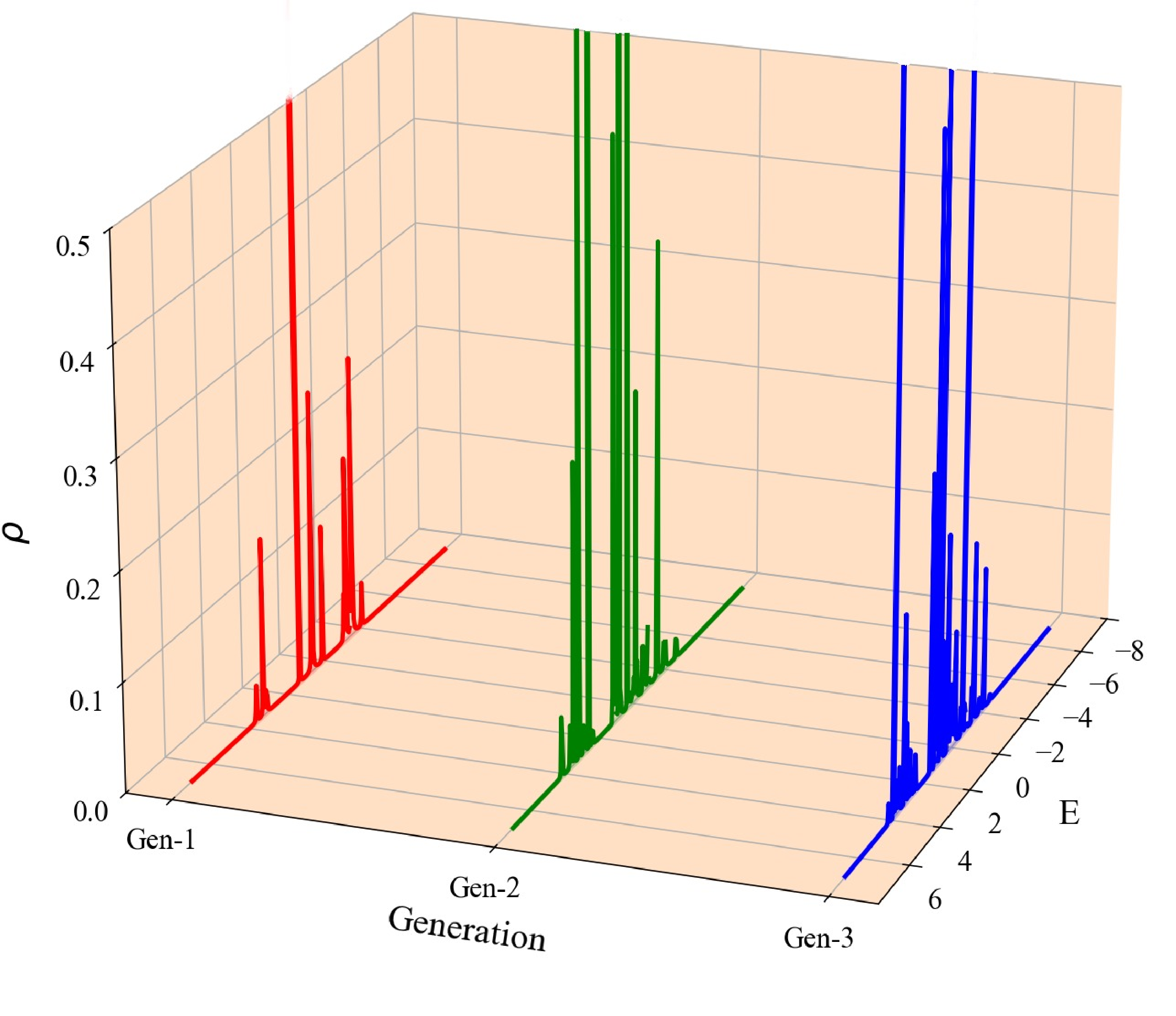}
		(c)\includegraphics[clip,width=0.40\textwidth]{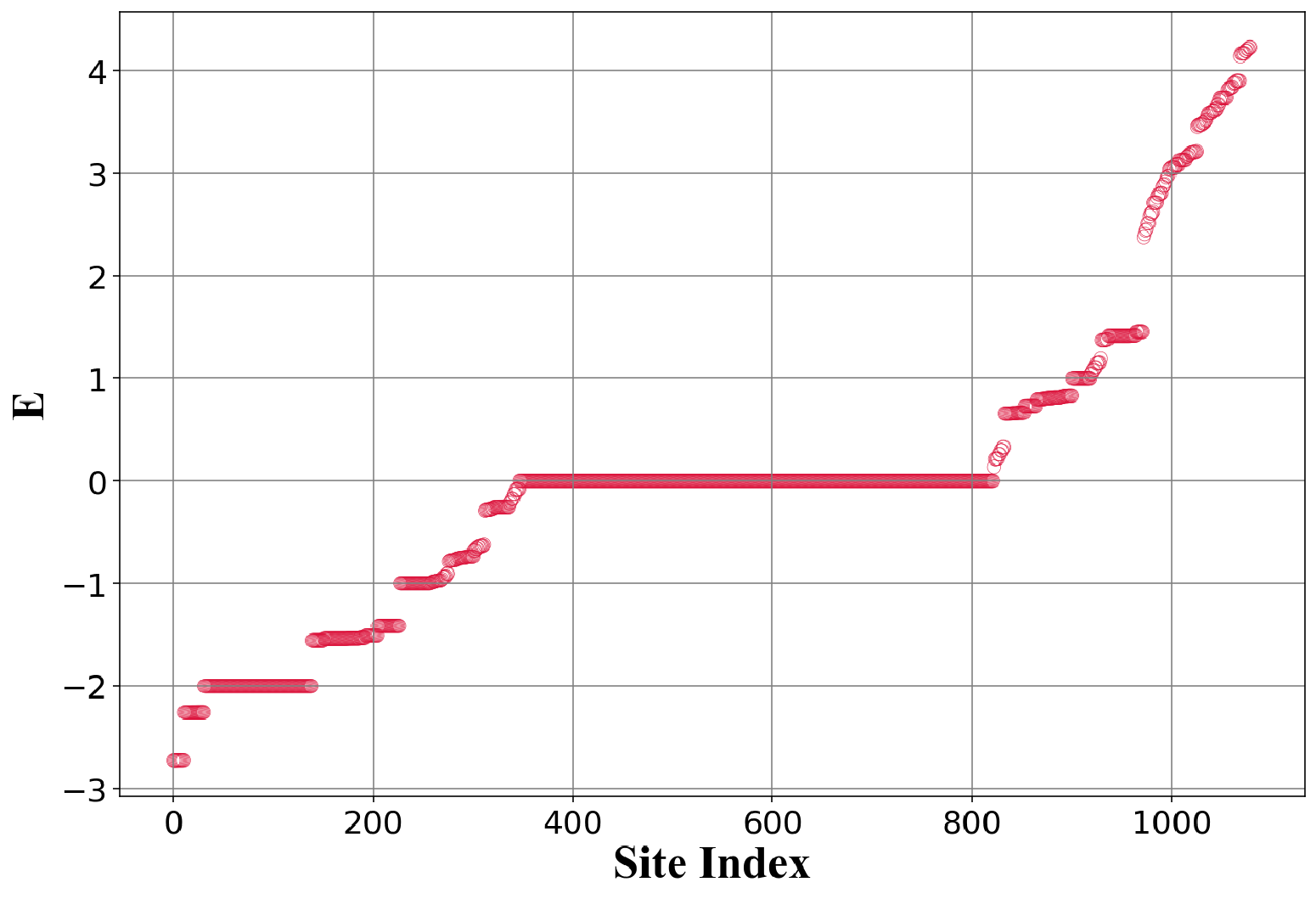}
		\caption{(a) Amplitude distribution for the localized state at $E = \epsilon -2t $ for first generation corral, (b) Distribution of localized states in the spectrum for successive three generations and (c) eigenvalue distribution of a third generation corral in absence of flux.}
		\label{ampdis}
	\end{figure}
	
With the initial description of the Hamiltonian, it is simple to check that in absence of flux, if we set, $E = \epsilon - 2t$, a consistent solution to the difference equation (Eq.\ref{dif}) can be achieved for which the amplitude distribution profile is pictorially shown in the Fig. \ref{ampdis}(a) for first Generation corral. As we observe that sites containing non-zero wave function amplitudes form the characteristic trapping island and one such island is effectively detached from the other by a special vertex where the amplitude of the wavefunction is zero. This typical profile does not allow the dynamics of the wave packet corresponding to energy eigenvalue $E = \epsilon - 2t$. Thus, the physical barrier formed due to destructive kind of quantum interference leads to frozen kinematics of the electron. The non-zero amplitudes(corresponding to $E = -2$) are concentrated over few atomic sites leading to the non-diffusivity of the associated wave packet. Moreover, one can exploit the self-similar pattern of the structure to obtain the hierarchy of localized modes. The level of hierarchy determines the areal span of the trapping cluster. The eigenstate resembles the spirit of molecular state \cite{skrk} localized due to wave interference resulted from the network. We expect a spectral singularity corresponding to this localized state because density of states $\rho \sim \int v_g^{-1} d^3k$. In the associated table.~\ref{tab:dos_values} we have checked the divergence with gradual decrement of the imaginary part $\delta$ added to the energy. As the generation of the corral incrases, the number of such spiky localized modes also increases as shown in Fig.~\ref{ampdis}(b). All these bound states are macroscopically degenerate and the degree of degeneracy is apparent from the Fig.~\ref{ampdis}(c).  

\begin{table}[h]
	\centering
	\begin{tabular}{c c c}
		\hline
		$E$ & $e_{\mathrm{im}}$ & $\mathrm{DOS}(E=-2)$ \\
		\hline
		$-2$ & $10^{-2}$ & $3.185\times10^{0}$ \\
		 & $10^{-3}$ & $3.182\times10^{1}$ \\
	 & $10^{-4}$ & $3.151\times10^{2}$ \\
		 & $10^{-5}$ & $1.591\times10^{3}$ \\
		\hline
	\end{tabular}
	\caption{Spectral divergence corresponding to $E = -2$}
	\label{tab:dos_values}
\end{table}

	\section{Impact of magnetic perturbation}
	\label{mag}

	\subsection{Flux dependent eigen spectrum}
\begin{figure}[ht]
	
	\includegraphics[clip,width=0.45\textwidth]{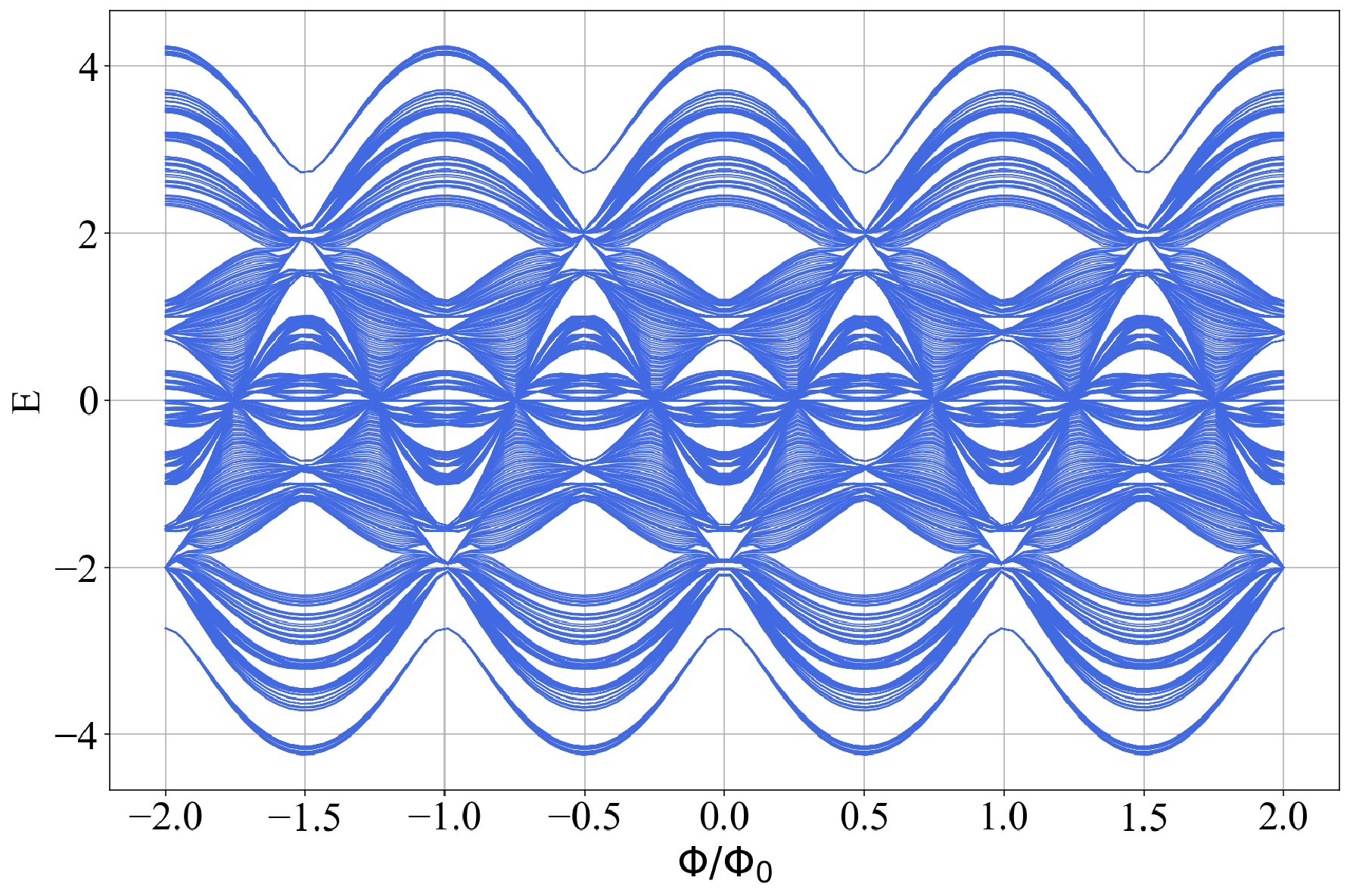}
	\caption{Flux dependent periodic eigenspectrum of a third generation quantum corral}
	\label{spectrum}
\end{figure}

All the relevant spectral signature including the transport characteristics can be found from the tight binding description of the system. To discuss the immediate impact of applying external magnetic perturbation over the single particle eigenstate and magneto-transport is the key objective of the analysis. Before going to the detailed demonstration, we shall try to get an overview of the flux dependent permissible eigen spectrum of the scale invariant corral structure. The straightforward diagonalization of the Hamiltonian matrix can reveal the flux-periodic landscape as shown in Fig.~\ref{spectrum}. The dimension of the matrix is essentially dependent on the level of hierarchy of the self similar fractal. We have cited the spectrum for the third generation fractal containing 1080 sites. The periodicity of the pattern is self explanatory and is reflected from the standard AB-phase via the off-diagonal entries of the Hamiltonian. The clustering of eigenvalues in presence of magnetic flux provides us the exotic spectral scenario where we see several interesting inter-twined band crossings and specifically flux dependent modification of band curvature. Hence one can have direct estimation of the mobility of the incoming excitation with respect to the applied flux, an external perturbation. This can be used to manipulate the imprisonment of electron in subtle way. The distribution of eigenvalues also depends remarkably on the generation of the aperiodic corral structure. The scale invariance of the pattern is apparent from the diagram. It is needless to say that with the gradual increment of generation of corral, the spectrum gets saturated and becomes indistinguishable from that of previous generation and we eventually get the spectral flavor of thermodynamic limit.

If we carefully observe the flux sensitive portrait, we see that when flux is set  $\Phi = \frac{1}{4}\Phi_0$, a centro-symmetric absolutely continuous resonant band appears. In case of self similar fractal object, because of absence of translational ordering, we do not expect diffusive modes to appear in the band spectrum. Instead, it should be full of localized states. But here we address that specific external perturbation can invite delocalization of single particle eigenstates and to the best of our knowledge this has not been addressed before in such quantum corral network. For $\Phi = \frac{1}{2}\Phi_0$, the resonant band disappears leading to the formation of bound modes again. Thus the caging of excitation is highly dependent on the magnetic flux. The phase coherence helps the prisoner to escape from being localized within the quantum prison. Flux sensitive band engineering and associated quantum interference destroy the Bloch-band and makes the prisoner trapped. External perturbation controlled periodic change of the localization-delocalization dynamics may inspire the experimentalists.

	\begin{figure*}[ht]
		\centering
		\includegraphics[clip,width=0.45\textwidth]{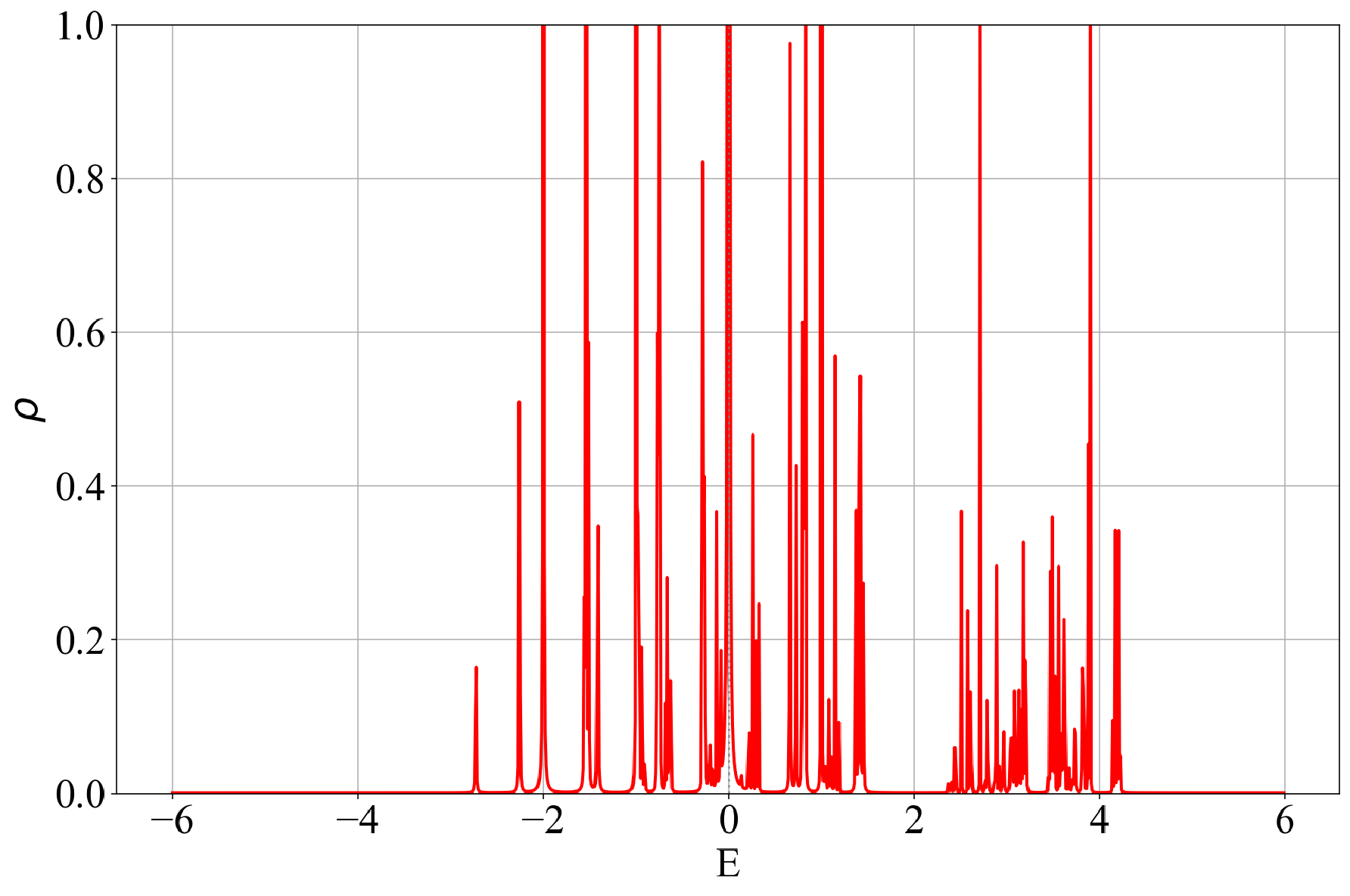}
		\includegraphics[clip,width=0.45\textwidth]{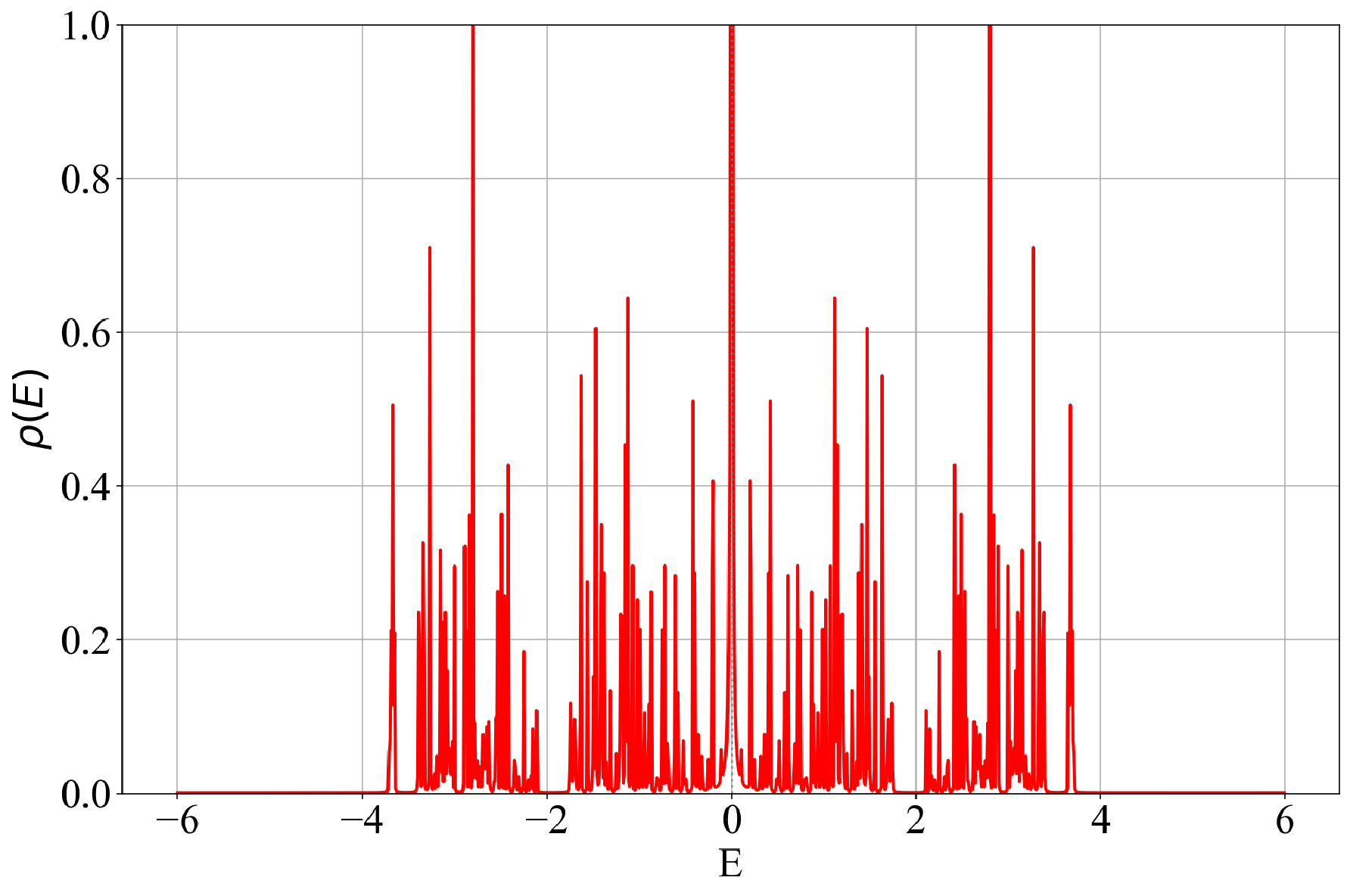}
		\includegraphics[clip,width=0.45\textwidth]{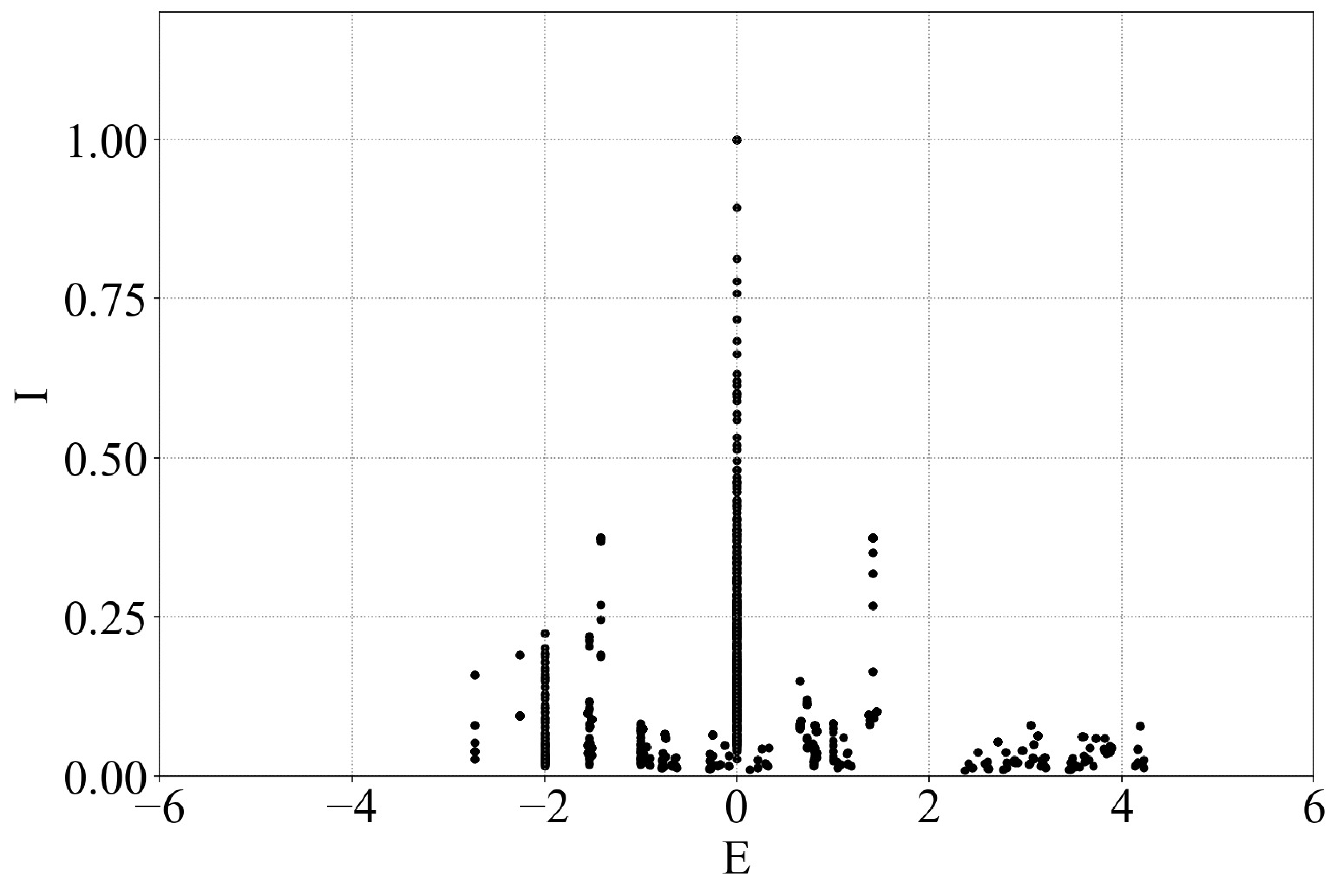}
		\includegraphics[clip,width=0.45\textwidth]{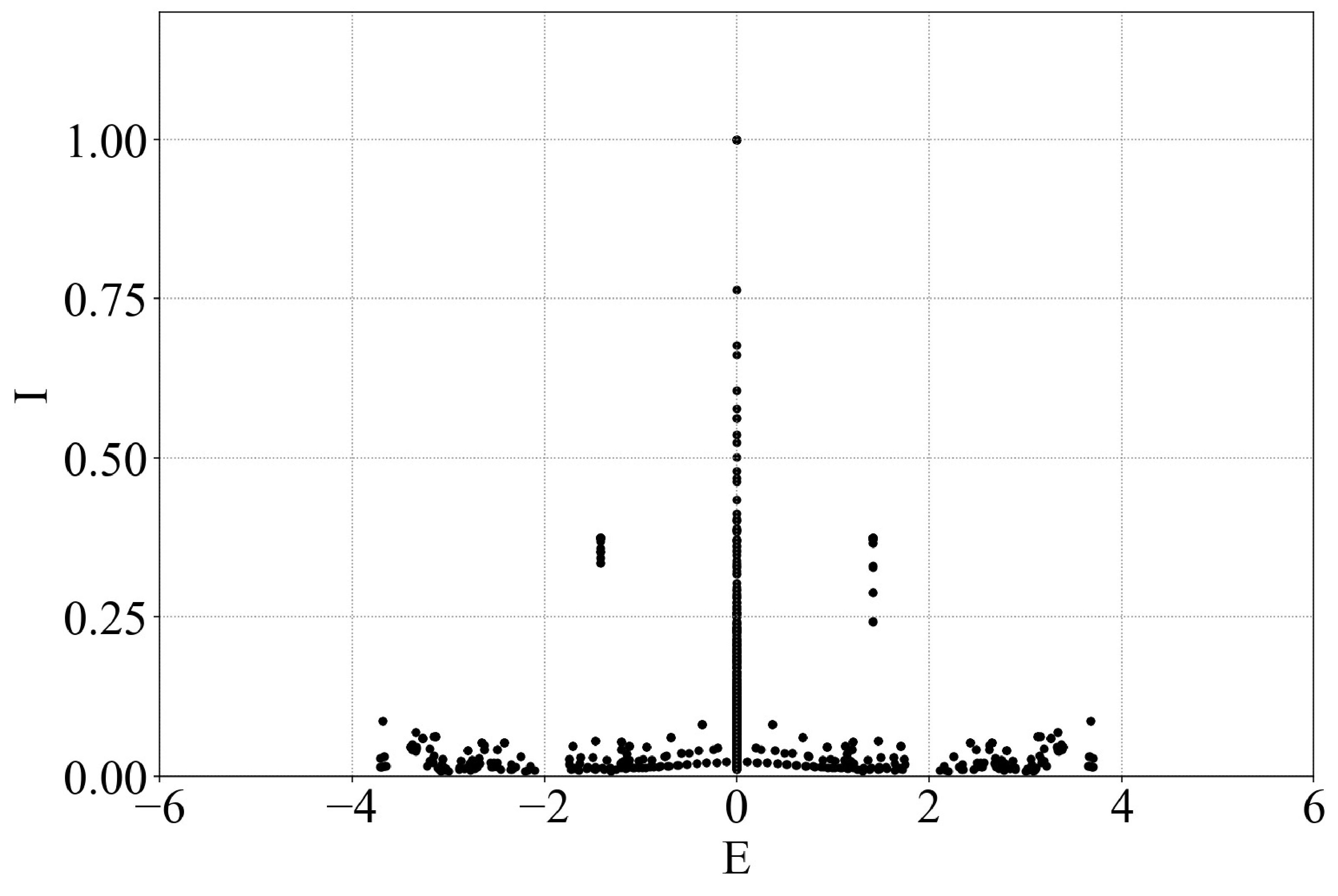}

		\caption{(Upper panel) Variation of electronic density of states with energy for (a) $\Phi = 0$ and (b) $\Phi = \frac{\Phi_0}{4}$. (Lower panel) Plot of inverse participation ratio against energy for (a) $\Phi = 0$ and (b) $\Phi = \frac{\Phi_0}{4}$.}
		\label{dos}
		
	\end{figure*}


\subsection{Density of states}
For precise knowledge about the single particle eigenstate, the preliminary step is to formulate the density of eigenstates (DOS) profile as a function of energy $E$ of the incoming projectile. The Green's function formalism helps to obtain DOS for the third generation (total number of atomic sites = 1080) corral structure using the following expression,
\begin{equation}
	\rho(E) = -\frac{1}{N\pi} \lim_{\eta \to 0} G_{00}[E + i\delta]
\end{equation}
The DOS pattern, as exhibited in Fig.~\ref{dos}, is checked, as permitted by the limit of accuracy, to be stable against decreasing the value of the imaginary part added to energy. The range of DOS is displayed within a value of unity to give prominence to the smaller peaks compared to the relatively larger ones. As we see that in absence of external flux ($\Phi = 0$), the fragmented and scanty appearance is very common for such kind of fractal entity. This is the foremost impact of structural aperiodicity.

A finite magnetic field, however, is competent enough to bring phenomenal changes into the spectrum even when we deal with a particular system with a pre-defined set of parameters. The second plot of Fig.~\ref{dos} is quite interesting. The spectrum presents very closely spaced zones of finite DOS. Particularly, the central absolutely continuous regime of resonant eigenfunctions around $E = 0$ is a prominent feature, as noticed. We have made a very fine scan around $E = 0$, each time reducing the energy interval to be scanned and diminishing the imaginary part $\delta$. Within the limit of machine accuracy, it is really tempting to conjecture the existence of a conducting regime around $E = 0$. For any energy belonging to the continuum, the system becomes highly transparent to the injected electron leading to the delocalization of eigenstates. The non-zero overlap of the wavefunctions between the nearest neighbouring sites for any energy within the continuum manifests the extendedness of the associated wavefunction. This feature is interesting to note and in distinct contrast to that obtained as we turn off the external field. The occurrence of a dense cluster of non-zero DOS has been checked carefully with other flux values ($\Phi < \frac{\Phi_0}{2}$). The magnetic flux thus can control the spectral response of the quantum corral structure in a comprehensive way. Magnetic flux can modify the nature of quantum interference, thereby decreasing the number of disallowed modes (gaps) and making the system much more accessible to the excitation. This indicates possibility of engineering the quantum states with the aid of external parameters.

	\subsection{Inverse participation ratio }
In this part, we have examined the variation of the inverse participation ratio (IPR) for the hierarchical structure as a counterpart of the study of electronic density of states. It is certainly a standard strategy that helps us to comment on the nature of the single particle eigenstate. The formal definition goes as the fourth power of the normalized wavefunction, i.e.,
\begin{equation}
	I \sim \sum_{j} \psi_j^4
\end{equation}
where the upper limit of the above summation depends on the level of hierarchy. This study essentially gives an idea about the cluster of atomic sites participating in any particular eigenfunction. Ideally, for a Bloch-like extended eigenstate, it takes zero value while it approaches unity for localization of the wave function. We have used this estimation of IPR as a first tool to corroborate the DOS results because it no longer assumes the exponential localization of the wave function. Hence this workout can give a general flavor of the spectrum. The results are shown in the lower panel of the Fig.~\ref{dos} for the third generation corral. The plot shows that for $\Phi = \frac{1}{4}\Phi_0$, we see an almost flat low-IPR sub-band around $E = 0$ leading to ballistic transmission profile. This central sub-band directly demands the existence of diffusive modes for this non-translationally invariant self similar model. The relatively high IPR modes tell about the imprisonment of the electron. Thus the plots show the dramatic change of localization length of the eigenstate with the variation of magnetic flux. A more rigorous approach is discussed in the next section to characterize the nature of the state.

	
	\section{Flux Driven extended states}
\label{flux}
\begin{figure}
	\includegraphics[clip,width=0.40\textwidth]{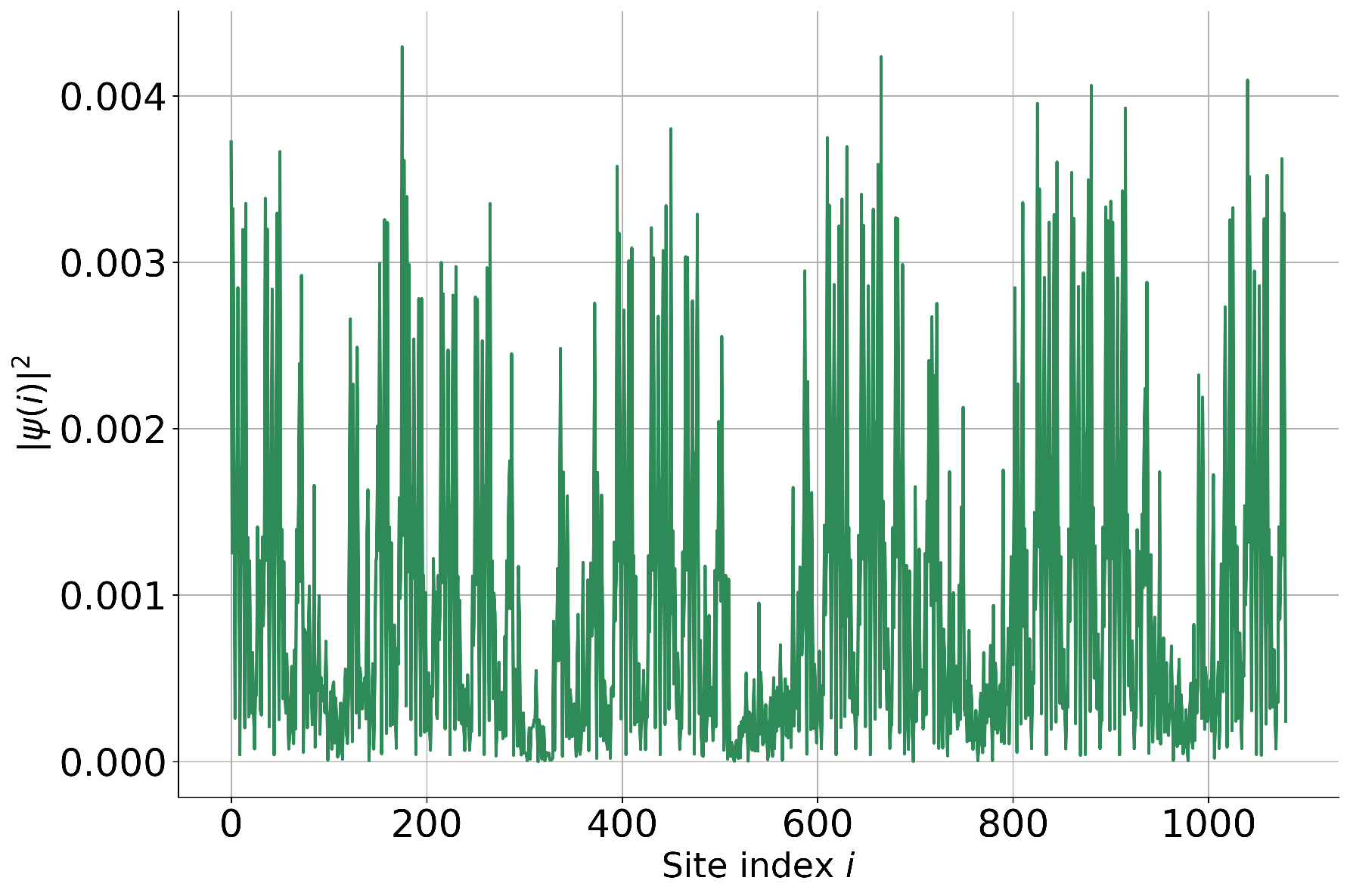}
	\caption{Oscillating $|\psi^2|$ vs. $n$ plot for $E = 1.35$ indicates the extendedness of the state.}

	\label{mospsi}
	
\end{figure}	
\subsection{General Remarks}
The escape of the prisoner in such a fractal substrate without any transnational order has always been an intriguing feature in the spectral study of disordered systems. To establish this, we have relied on a careful numerical workout. The straightforward and minimal check for the extendedness is the plot of ${|\psi|^2}$ against the number of sites present in any finite generation of the fractal. In Fig.~\ref{mospsi}, we have plotted it for a suitable energy $(E = 1.35)$. First, the central absolutely continuous sub-band in the DOS spectrum is found to be very stable as the imaginary part added to the energy is decreased chronologically from $10^{-3} \text{ to } 10^{-6}$. A very fine scan over the points in this interval reveals that for any energy eigenvalue we hit upon quite randomly in this range, non-vanishing ${\psi^2}$ indicates the diffusiveness of the corresponding eigen state obtained at a particular level of hierarchy. The same nature follows for the other nature of the continuum. The flux-induced phase coherence thus emphasizes the existence of resonant states. In the subsequent discussion we will try to get the overview of the response of the system in a different way.

	
	\subsection{Transport Characteristics}
	
		\begin{figure}[ht]
		\centering 
		\includegraphics[clip,width=0.40\textwidth]{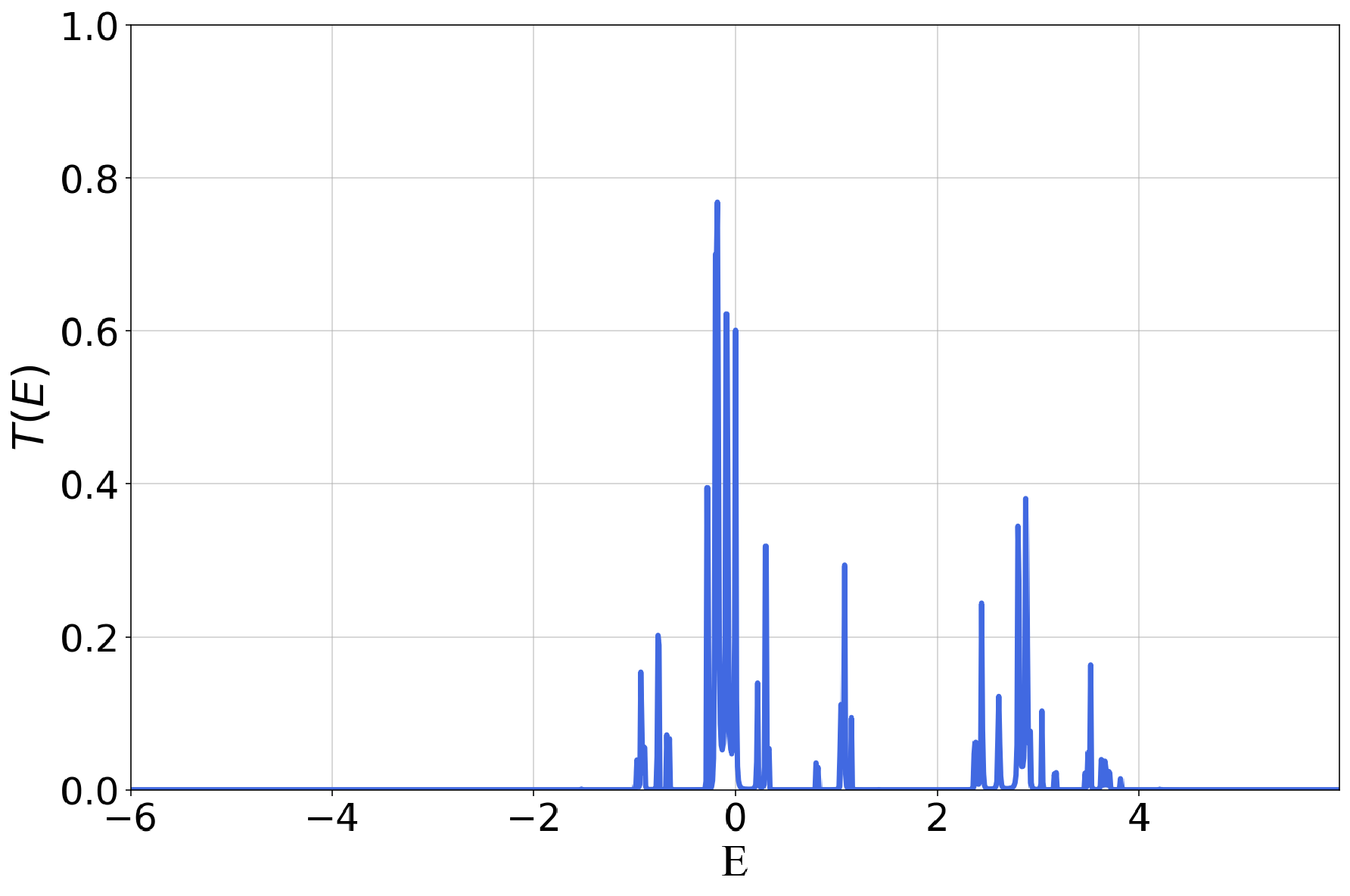}
		\includegraphics[clip,width=0.40\textwidth]{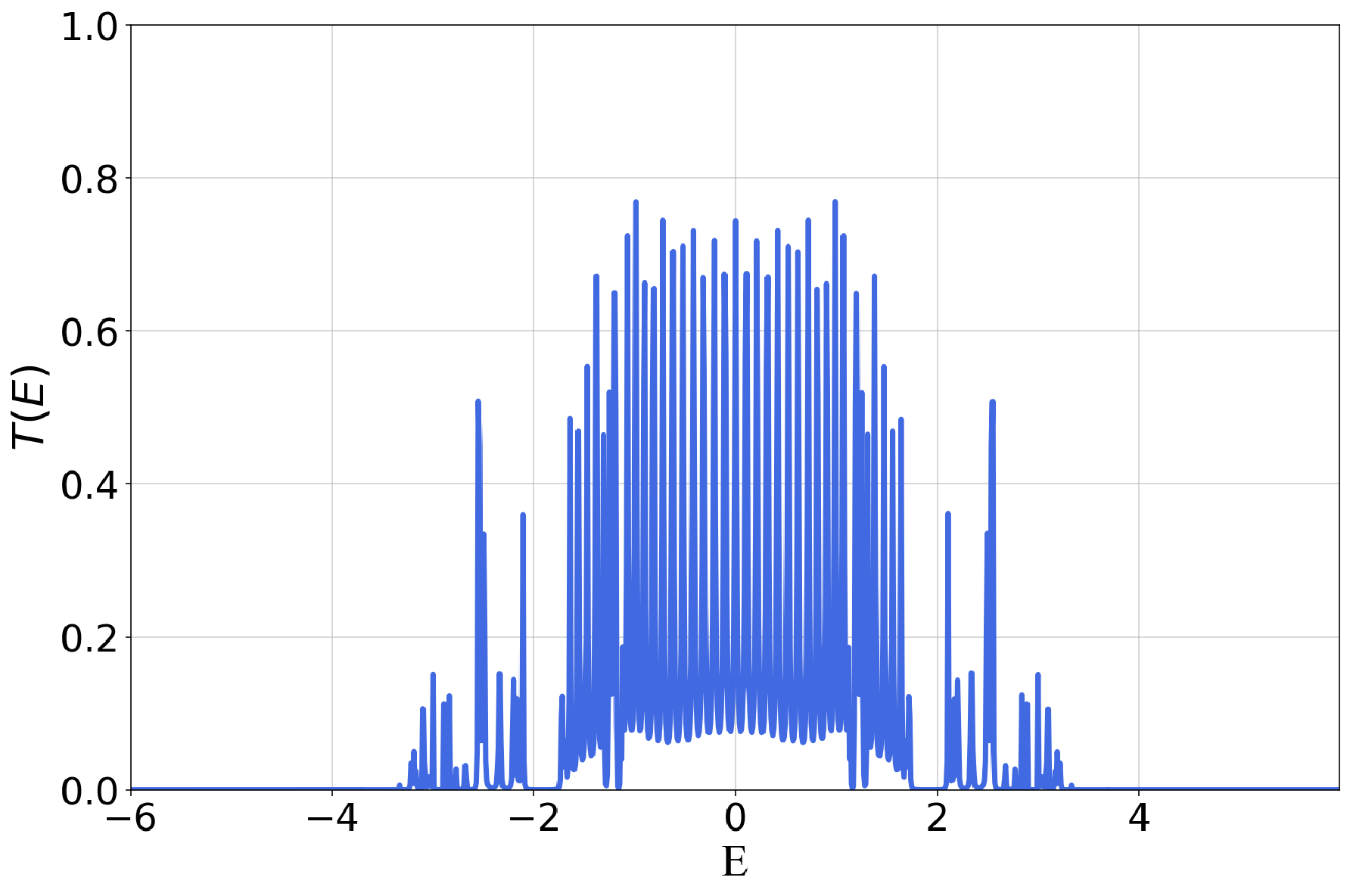}

		\caption{Graphical representation of transmission probability as a function of energy for (a) $\Phi = 0$ and (b) $\Phi = \frac{\Phi_0}{4}$.}
		\label{transport}
	\end{figure}
	
To validate our finding, we now study the variation of quantum transport across a corral structure of any definite size. The essential methodology consists in positioning the desired fractal geometry in between two pairs of (named as source and drain respectively) semi-infinite perfectly ordered leads. The leads may be demonstrated in the tight binding language by a uniform site energy $\epsilon_0$ and overlap parameter $t_0$. With the initial description of the arrangement the transmission probability \cite{Mujica1994} can be calculated using the following expression,

\begin{equation}
	T_{mn} = 4 \Delta_m \Delta_n \left| G_M \right|^2.
\end{equation}
Here $\Gamma_m$ and $\Gamma_n$ denote the connection of the system
to the $m^{\text{th}}$ and $n^{\text{th}}$ lead respectively, and 
$G_M^{r}$ and $G_M^{a}$ are the retarded and advanced Green's 
functions of the system, respectively. As the coupling matrix 
$H_{mM}$ is non-zero only for the adjacent points.

The choice of numerical values of the lead parameters and the placement of lead have been made judiciously to capture the response of the system over the entire range of energy eigenvalue. The results are plotted for third generation of the corral structure sequentially in the Fig.~\ref{transport} both in absence  and presence of magnetic perturbation.

As it is reflected from the DOS and IPR plots, when the magnetic flux is not switched on, the transport behavior contains some isolated peaks (or very narrow sub-bands) consistent with the DOS spectrum. In absence of flux the overall tranmittivity of the corral is very low, as expected. With increasing the hierarchy of the corral, it incidentally turns out to be even poorly transmitting, as expected. Interesting changes however start showing up as we turn on the external flux $\Phi$. With the introduction of nominal perturbation, regimes of finite transmission increase in number. These windows of appreciable transmittance join \textit{hand in hand}
 to form a Bloch-band with the gradual increment of $\Phi$. As it is shown in the Fig 5 that for $\Phi = \frac{1}{4}\Phi_0$, the spectrum becomes notable for a thick population of resonant eigen functions at and around $E = 0$. Prominent broadening of the central conducting band tells us that magnetic flux makes the fractal more accessible to the incoming electron. This picture is in sharp contrast to that in absence of flux. We have gradually increased the flux value and seen that the energy span of diffusive modes enhances with flux and it becomes maximum at $\Phi = \frac{1}{4}\Phi_0$. The symmetric spectrum gets shrunk as we move towards half flux quantum. The change of width of the transmitting windows cites a periodic variation with the flux and this flux periodicity is self-explanatory.

	\section{Robustness of the extended states} 
	\label{rob}
	\subsection{In presence of diagonal disorder}
	
			\begin{figure}[ht]
		\centering 
		(a)\includegraphics[clip,width=0.45\textwidth]{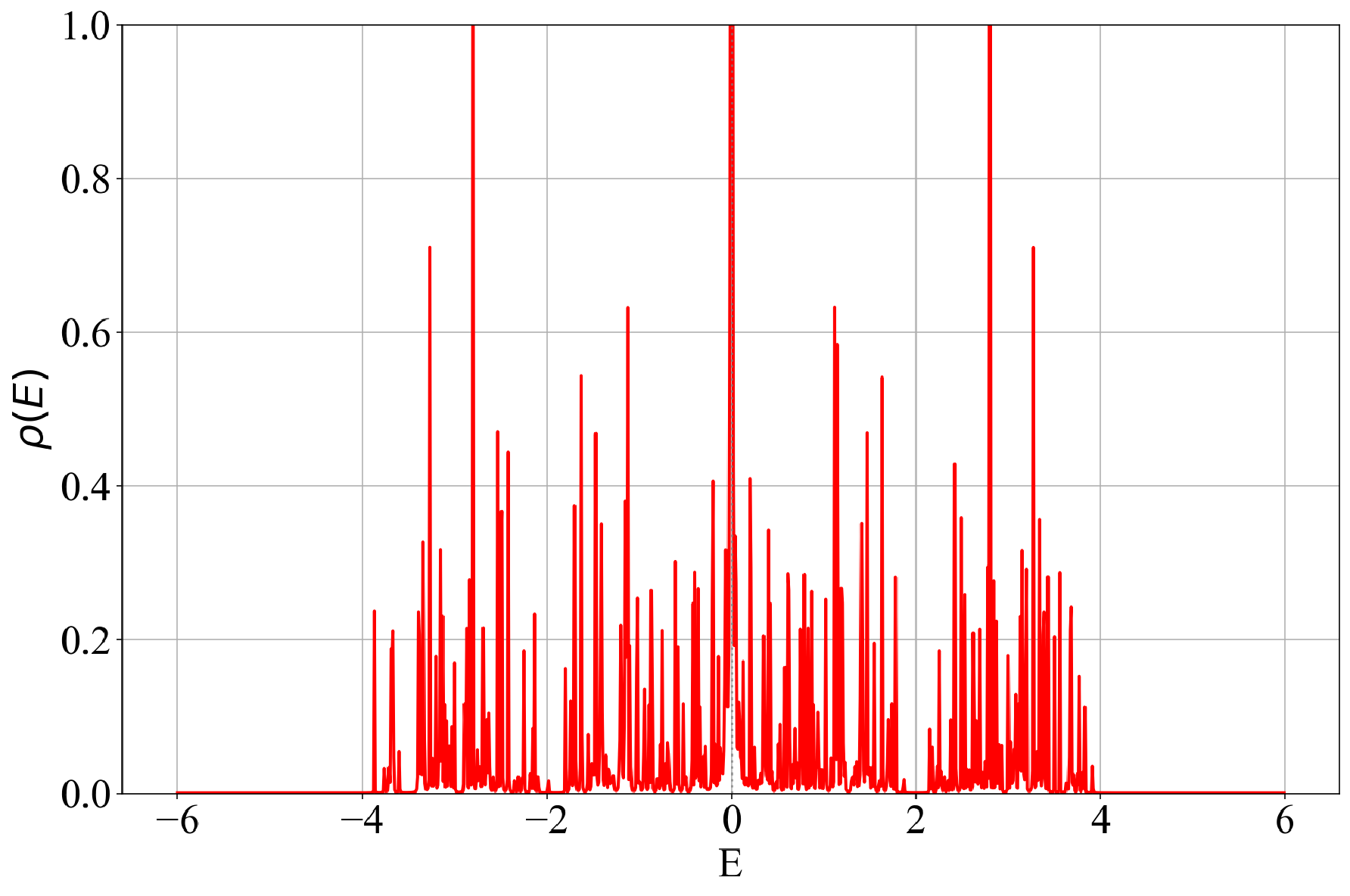}
		(b)\includegraphics[clip,width=0.45\textwidth]{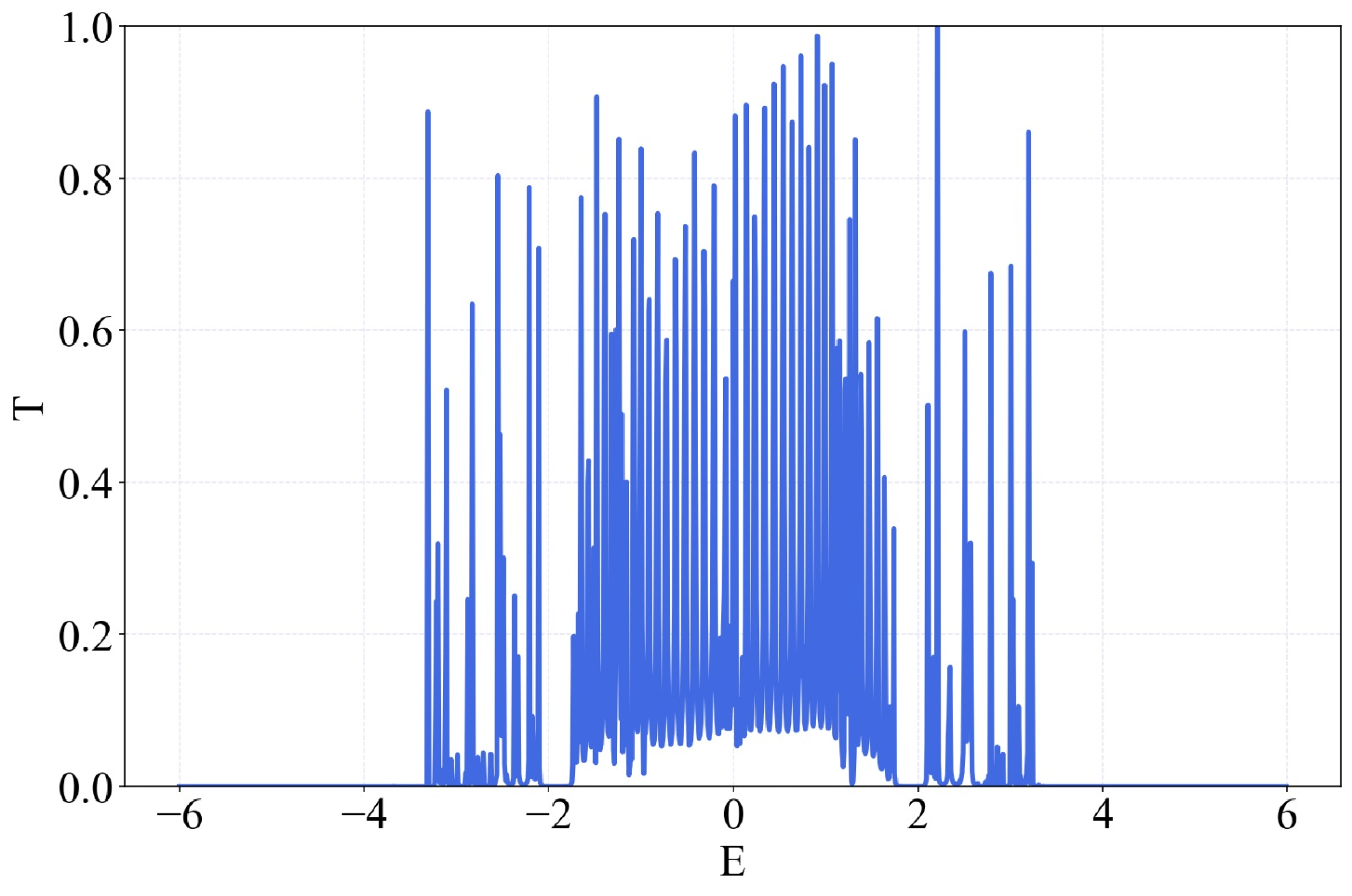}
		
		\caption{Variation of (a) density of states and (b) transmission coefficient with energy for $\Phi = \frac{\Phi_0}{4}$. The strength of disorder is taken as $w = 0.5$.}
		\label{disdos}

	\end{figure}

	At this point we will try to address the sensitivity of flux induced escape from localization of single particle eigenstates in presence of diagonal disorder. This aspect may be examined with the inspection of DOS spectrum in a disordered environment at non zero flux values. Here, we have taken a random variation of on-site potential $\epsilon_j$ and calculated the eigenspectrum in presence of magnetic perturbation. The result we have plotted in Fig.~\ref{disdos} is for third generation corral. We  observe that resonant window that appears in presence of non zero flux in such fractal network remains nearly unperturbed even in this randomly disordered environment. The potential arrangement follows a distribution, viz., 
	$\epsilon_j \in \left[-\frac{w}{2}, \frac{w}{2}\right]$, where $w$ denotes the strength of disorder. We have thoroughly checked the sustainability of the central Bloch-like sub-band for low to high strength of disorder. We have taken $w = 0.1$ to $1$ (measured in unit of \textit{t}). We can comment in a conclusive way that the resonant window becomes undisturbed against the potential disorder for $w = 0.5$. This has been justified by the workout of transport results as shown. The extremely ballistic(T) clearly indicates that the flux sensible escape of the prisoner is much more dominating. For large strength of disorder the bands may subdivided into parts but the some diffuse modes with relatively large localization length may still be present there. The central spike becomes detached from the continuum and it can highlight its localization character for $w$ exceeding $0.5$. The interesting as well as challenging spectral competition inspires the us to check and report the perturbation manipulative feature in this manner.

			\begin{figure}[ht]
		\centering 
		(a)\includegraphics[clip,width=0.30\textwidth]{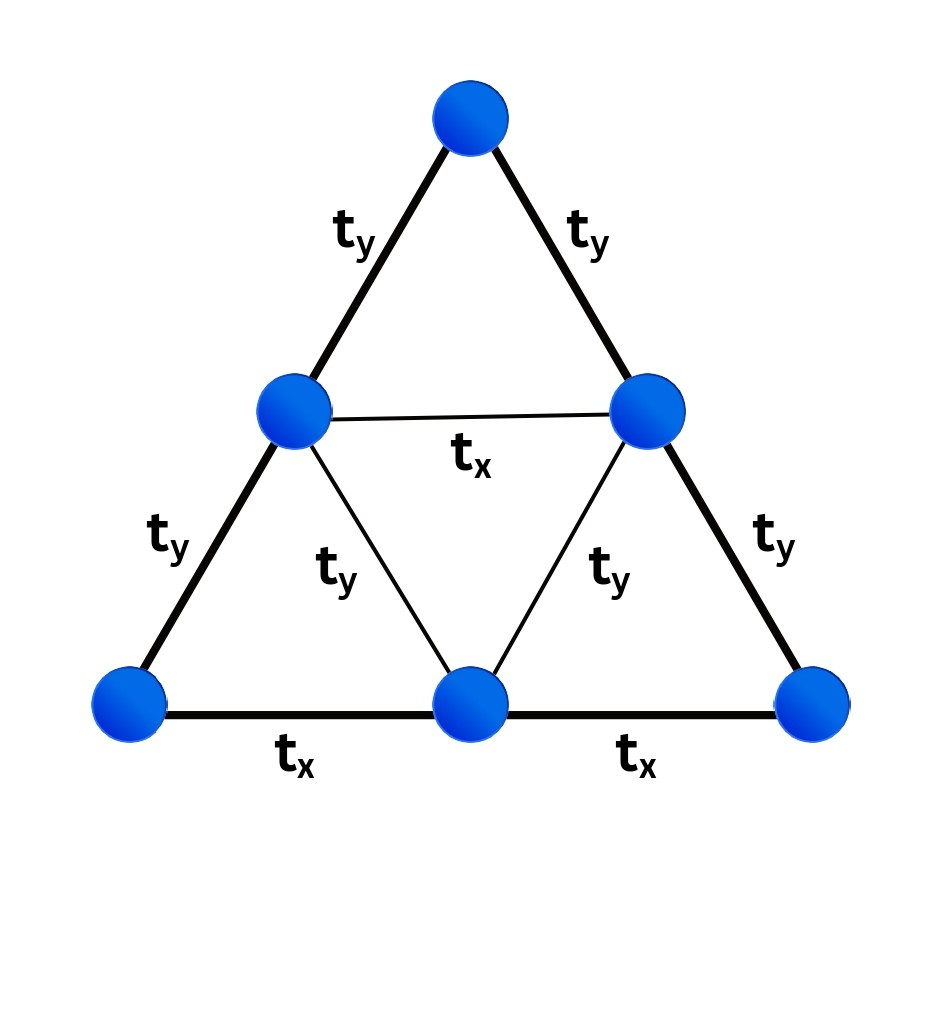}
		(b)\includegraphics[clip,width=0.45\textwidth]{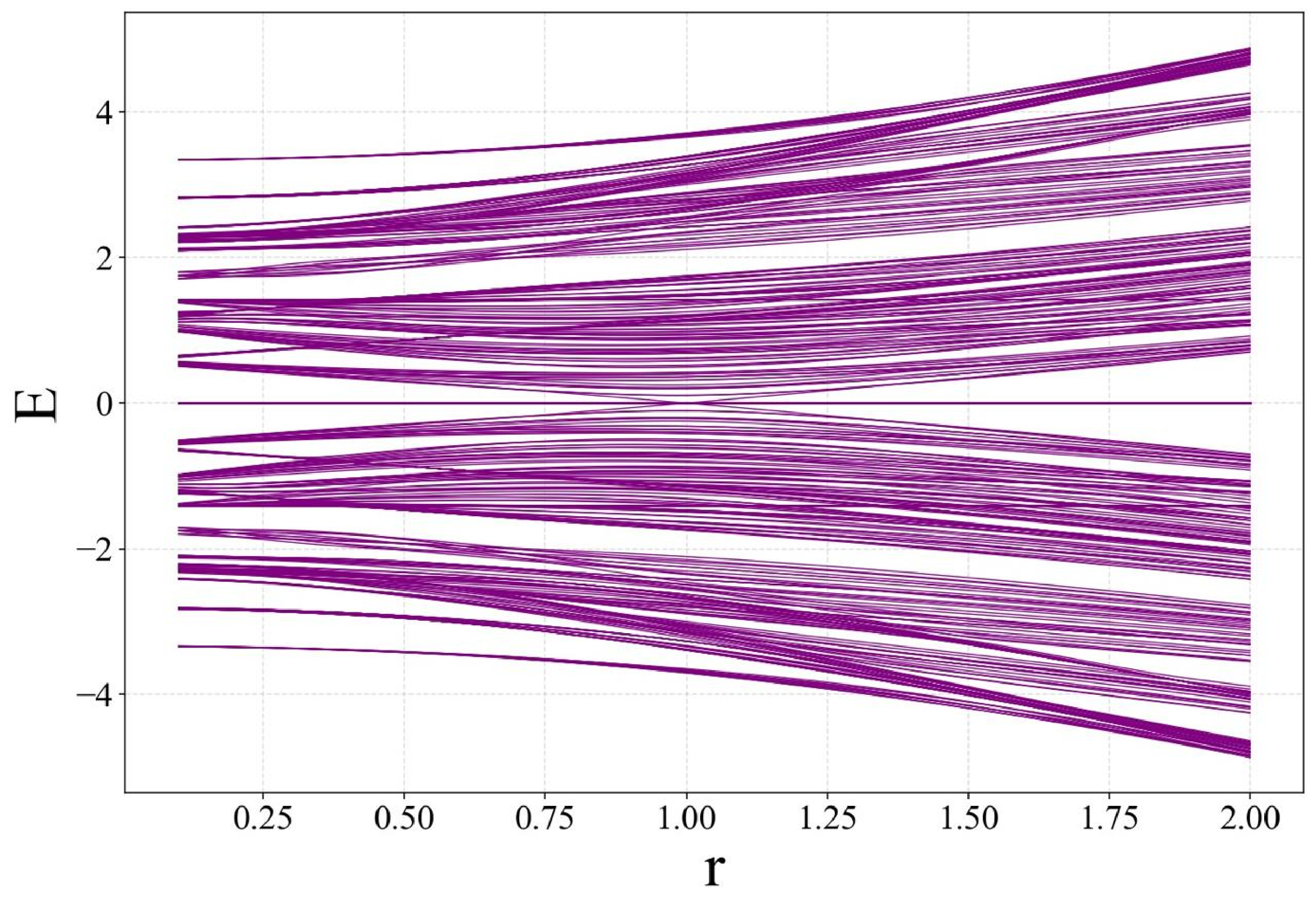}

	
		\caption{(a) Schematic view of anisotropic corral and (b) pictorial representation of eigenvalue spectrum with off-diagonal anisotropy index $r$.}
		
		\label{anisohop}
	\end{figure}

	\subsection{In presence of off-diagonal anisotrophy}

Flux induced appearance of resonant modes prompts us to check the sustainability in further reasonable approach. In the previous discussion, we have seen that random potential arrangement cannot disturb the continuum band considerably. Now, we introduce an anisotropy in the corral with minimal perturbation in the kinetic parameter of the Hamiltonian and check the possibility of generation of conduction band. We define two types of hopping integrals $t_x$ and $t_y$ as shown in Fig.~\ref{anisohop} and play with the relative strength of these two, i.e., $\frac{t_x}{t_y} = r$ (say). In Fig.~7 we have cited the eigenvalue spectrum with respect to the ratio \textit{r}. We see that for extremely low range of $r$ there are few narrow sub-bands and few localized states. With sequential increase in $r$, the possibility of getting more number of quantum path to transit leads to the formation of resonant bands. These resonant bands merge with each other to form a central continuum. The gap around $E = 0$ tends to vanish as $r$ goes towards unity. For $r = 1$, the central localization. states fall within the continuum and lose their localization character. Beyond $r = 1$, the gap again opens up and fragmentation of the bands occurs for larger values of $r$. But existence of extended states is however not ruled out in presence of off-diagonal anisotropy. In conclusion, we can say that this kinetic mismatch index can control the response of the flux which will be discussed later.

\section{Multifractal Analysis}
\label{mfan}
	\begin{figure}[ht]
	\centering 
	(a)\includegraphics[clip,width=0.40\textwidth]{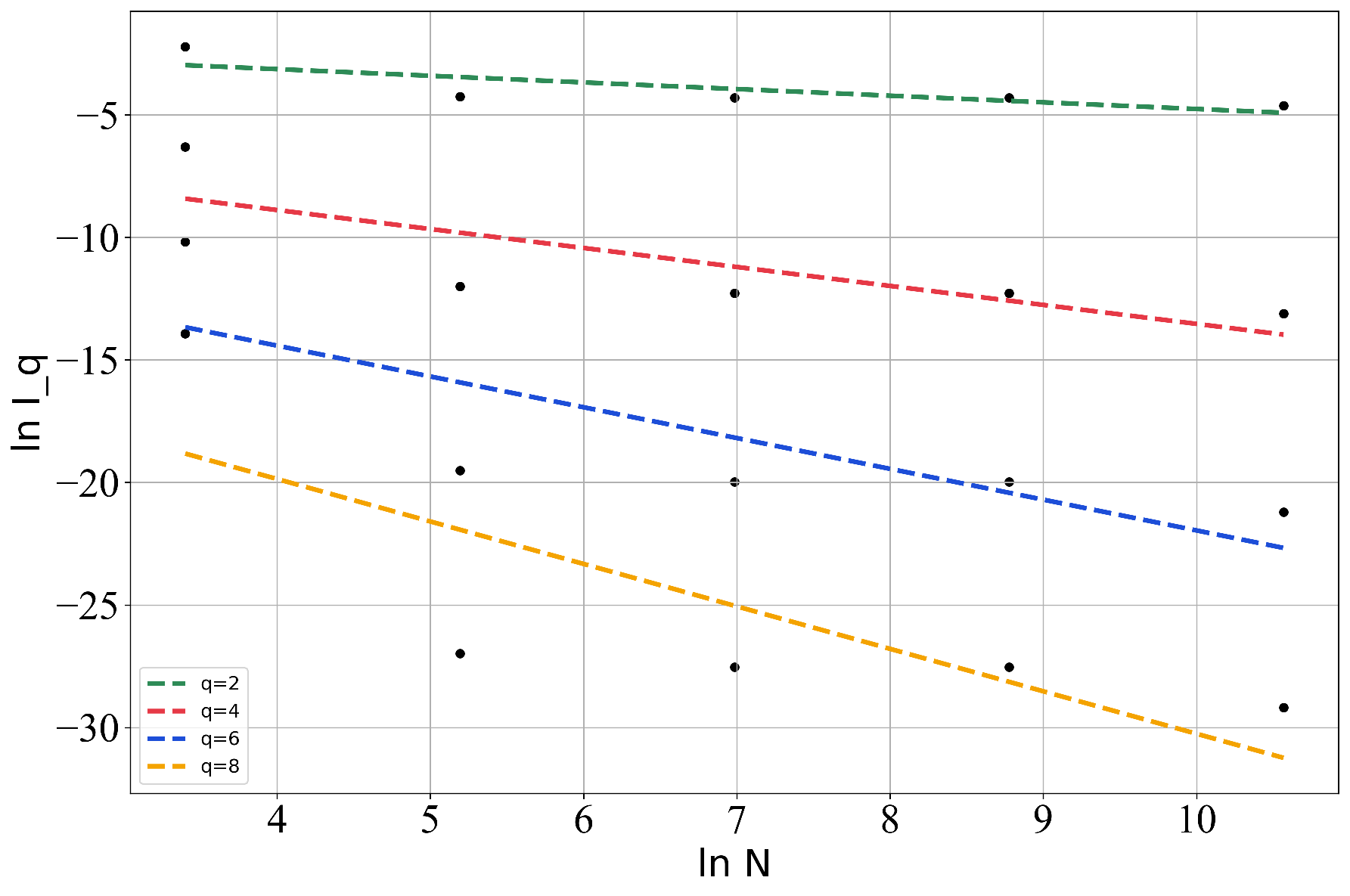}
	(b)\includegraphics[clip,width=0.40\textwidth]{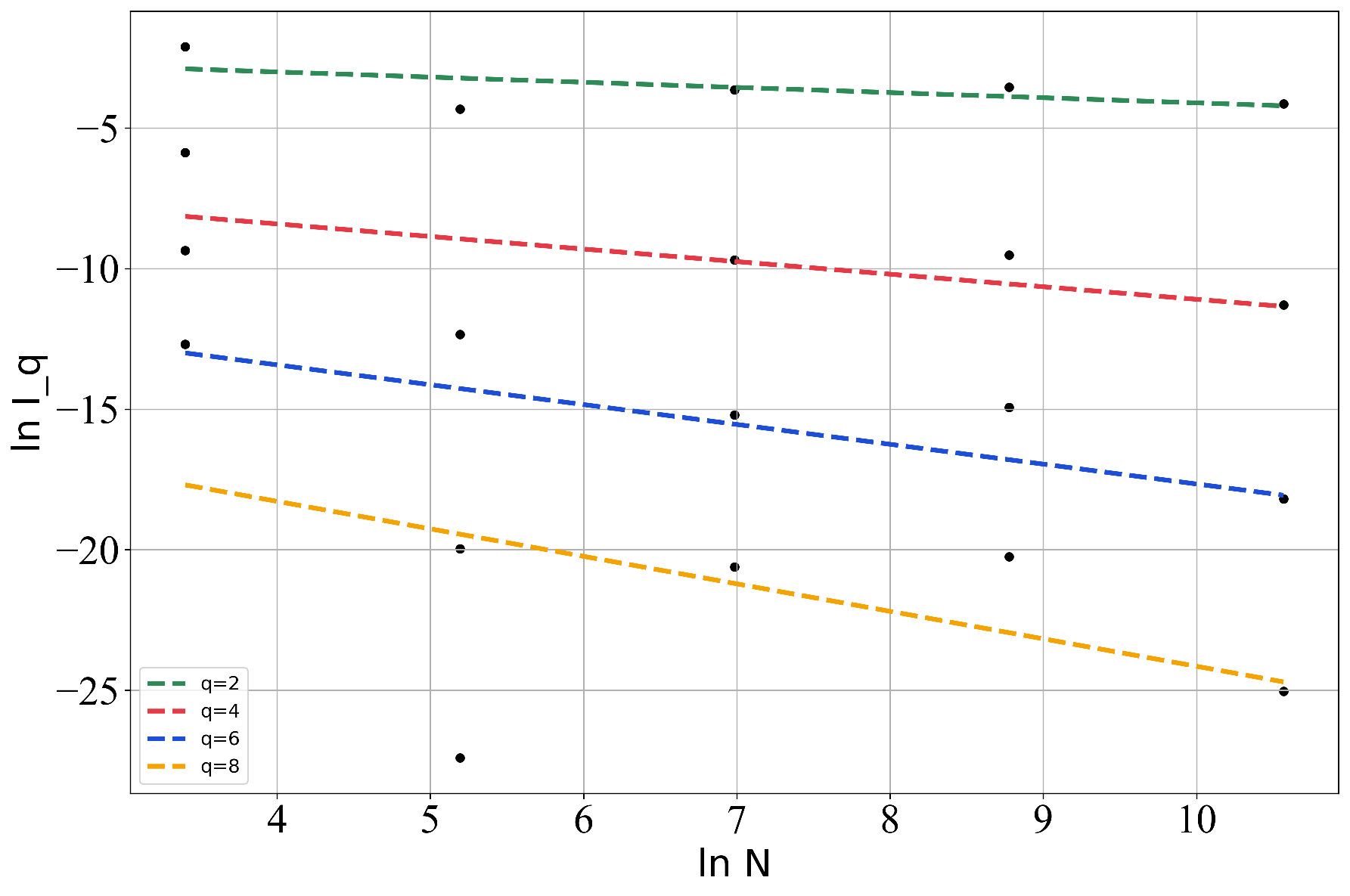}
	(c)\includegraphics[clip,width=0.40\textwidth]{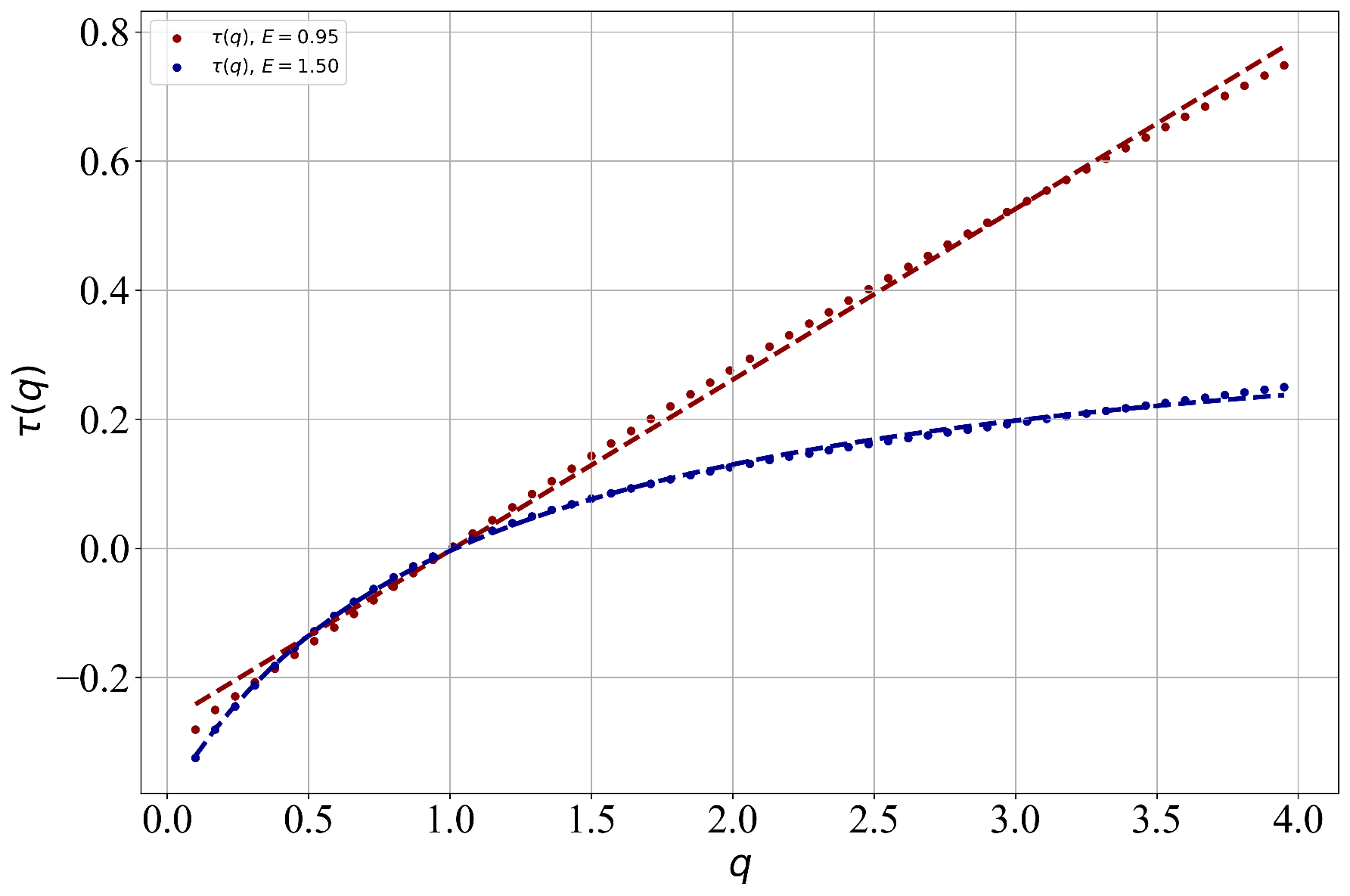}
	\caption{Representation of the $\ln I_q$ vs. $q$ plots for (a)$E = 0.95$, (b) $E = 1.5$ and (c) multifractality spectrum.}

	\label{mfa}
\end{figure}
	
We first focus on the variation of IPR of the entire spectrum against the energy $E$ of the electron. This plot is done for magnetic flux $\Phi = \frac{1}{4}\Phi_0$. As we move through the central subband, on either side of band centre, we see that IPR shows a dramatic jump around $E = \pm 1.414$, where the transmission is sufficiently low. This interesting change in the value of IPR (hence the localization length) becomes prominent as we increase the hierarchy of the corral. The absolutely continuous subband up to $E = \pm 1.414$ having low IPR values contains a closely spaced dense distribution of delocalized eigenfunctions. The extendedness of the wavefunctions has been confirmed through the transport profile, as mentioned in the previous discussion. At this point, we observe that high IPR modes differ almost by an order of magnitude compared to the low IPR states. This argument in the value of IPR is a first clear indication of existence of different quantum phases in the spectrum. To validate the nature of the single-particle states, a more meticulous approach has been taken in the subsequent discussion.

Multifractality of the wavefunctions, narrating their strong fluctuations at criticality, is a significant feature of the Anderson transitions. Unlike monofractals, multifractal networks are specified by a continuous set of exponents illustrating the scaling of moments of some probability distribution. For our case, the probability measure is $|\psi(r)|^2$. Now, we can define generalized IPR (GIPR) $I_q$ as the moments of eigenstate intensities~\cite{fe,hh},
\begin{equation}
	I_q = \int |\psi(r)|^{2q} \, d^d r
\end{equation}
At criticality, we expect a power-law variation of GIPR $I_q$ with the dimension of the Hamiltonian $N$, $I_q \sim N^{-\tau(q)}$. Here, $\tau(q)$ describes a continuous set of exponents which characterize a multifractal system. With this nomenclature, we can write $\tau(q)$ as $\tau(q) = (q-1)D_q$, where $D_q$ is known as the generalized fractal dimension. It is standard to be noted that the critical exponent $\tau(q)$ (a) is linear with $D_q = d$ for conductors, (b) becomes flat in localized systems with $D_q = 0$, and (c) is a nonlinear variation of $q$ at critical points.

As we see that, the spectrum exhibits two different regions of phases when we set the magnetic flux $\Phi = \frac{1}{4}\Phi_0$. One can reveal the character of the underlying quantum phases with the help of standard multifractal analysis. This formulation has been satisfactorily applied to several disordered and quasiperiodic lattice models ~\cite{lj,cg,hih}. With quartic flux, we select two eigenvalues from the spectrum $E = 0.95$ and $1.5$ from the low and high IPR regions respectively. The careful selection is made keeping the boundary at $E = 1.43$ in mind.

In Fig.~\ref{mfa} we have also cited the variation of generalized inverse participation (GIPR) $I_q$ against the system size for both the metallic state and critical state. The difference of the plots highlights the distinction between two phases. From these plots we also evaluate the multifractal exponent $\tau(q)$. For $E = 0.95$, the linear fashion of $\tau(q)$--$q$ plot unfolds the metallicity. While for $E = 1.5$, $\tau(q)$ varies nontrivially with $q$ and deviates more from the linearity as $q$ increases. This is an evident signature of multifractality. A nonlinear fit represents $\tau(q)$ changes as $\tau(q) \approx \frac{-8.5}{q + 5.3} + 1.4$, which readily explains the nonlinear variation becoming prominent for larger $q$. This workout finally demands that we can conclude the possibility of a single-particle mobility edge which is a domain wall specified by a certain energy separating the insulating phase from the conducting phase.

Before ending this discussion, we should mention that magnetic flux plays a severe role in the manipulation of the overall spectrum. Flux-sensitive extended states remain at the central part of the continuous subband but as our system has no translational ordering, reflection of the fractal network is observed as one looks away from the band centre. Coexistence of flux-induced resonant states and critical states prompts us to speculate that one should expect a transition between these two phases and this is the consequence of the interesting spectral competition between the structural aperiodicity and flux-controlled change in the kinematics of the electron.


	\section{Study of Persistent Current}
	\label{persi}

		\begin{figure}[ht]
	\centering 
	(a)\includegraphics[clip,width=0.44\textwidth]{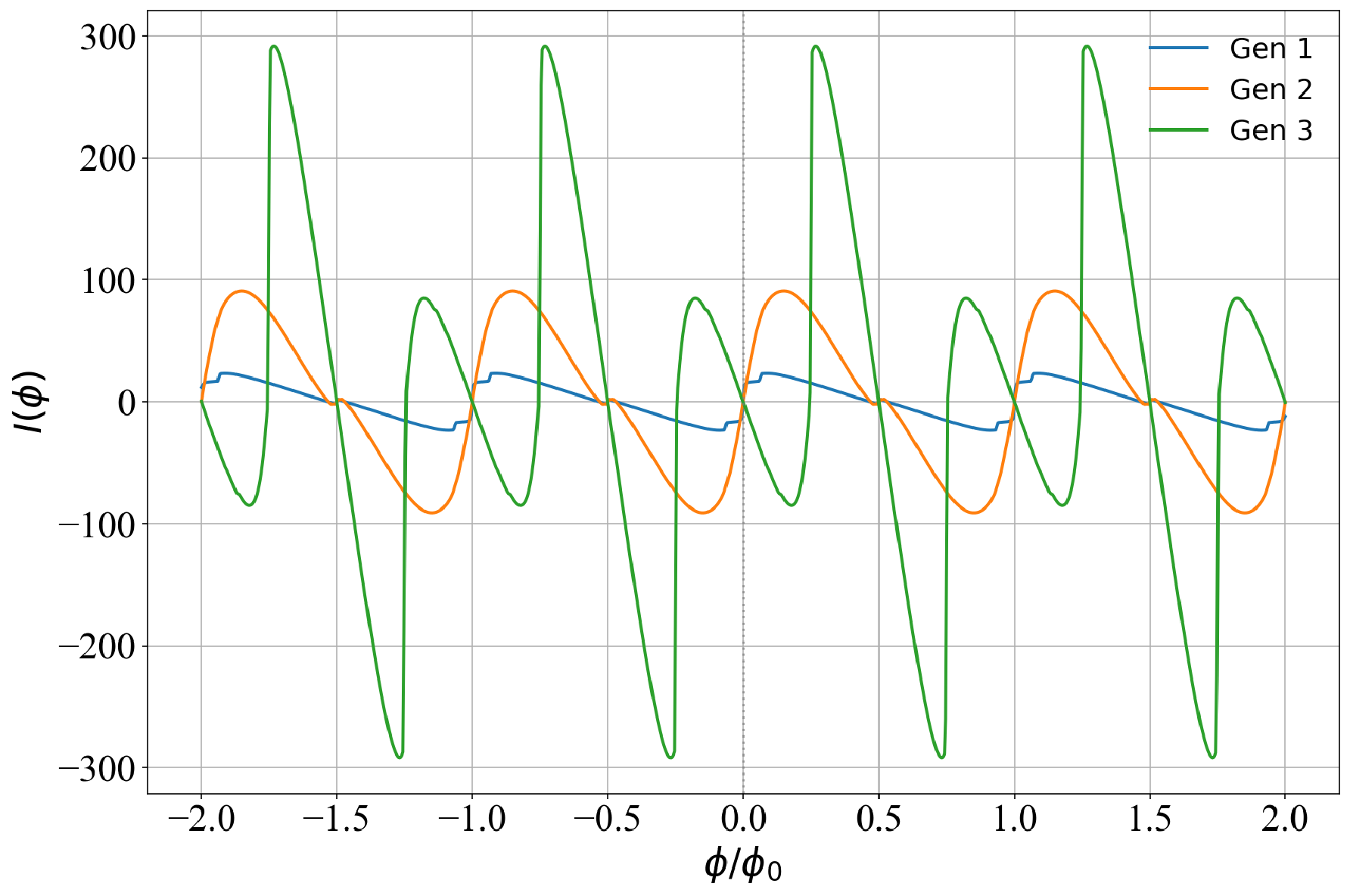}
	(b)\includegraphics[clip,width=0.44\textwidth]{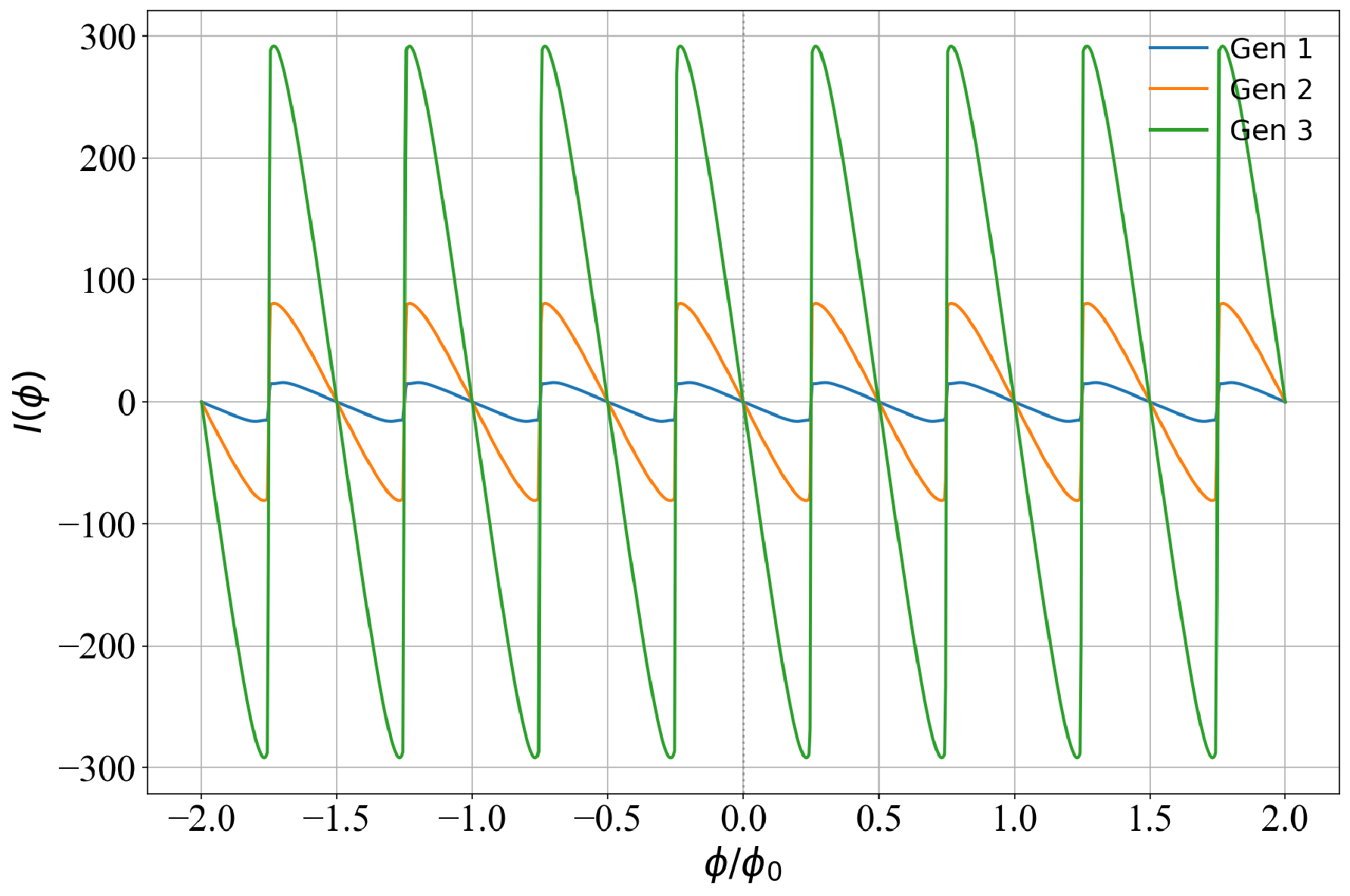}
	(c)\includegraphics[clip,width=0.44\textwidth]{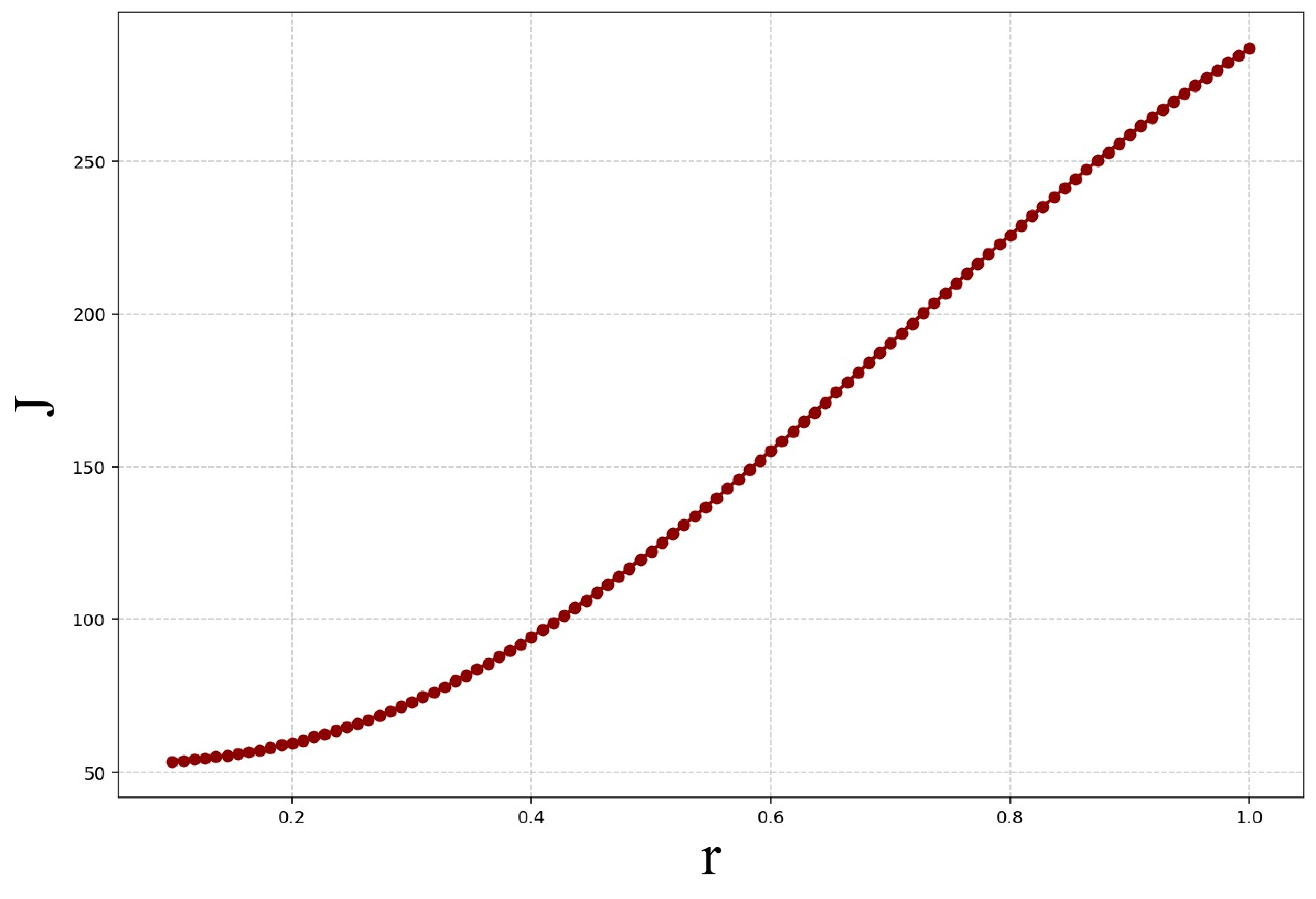}

	\caption{Variation of persistent current with flux for (a) $\alpha = \frac{1}{4}$ (b) $\alpha = \frac{1}{2}$ respectively. (c) Tunability of the magnitude of the current with the off-diagonal anisotropy.  }
	\label{persist}
\end{figure}	

From the previous discussion we can easily come to the conclusion that quantum interference happening in each elementary triangular loop is the essential thing which ultimately determines the dynamics of the prisoner (electron). The phase associated with electron wavefunction depends on the flux leading to a dissipationless current, oscillating with flux, named as persistent current ~\cite{gef}. Hence, for the sake of completeness of the spectral analysis, we have examined the flux sensitive oscillatory fashion for our self-similar corral substrate. It is needless to mention that this current is no longer related to that found in a superconducting element. This idea was first projected for a single mesoscopic conducting loop but eventually that was tested for multi-loop fractal geometries~\cite {pc}. The persistent current with respect to the magnetic flux $\Phi$ can be calculated using the standard relation, viz.,
\begin{equation}
	J(\Phi) = -A \sum _{j} \frac{\partial E_j}{\partial \Phi}
\end{equation}
Here, $A$ is a constant which is taken unity in our evaluation.
At this point, we now define a filling factor $\alpha$ that basically tells us about the proportion of the total energy levels filled. Fig.~\ref{persist} illustrates the variation of persistent current with flux $\Phi$ for three successive generations of the corral. We have shown the plots for $\alpha=1/4$ and $1/2$ respectively. The flux periodicity is inherent from the eigenspectrum. The increase in transparency with flux to the incoming electron for quarter flux makes the current maximum at that flux and the spectral collapse at half flux quantum is also reflected in the variation unfolding the concept of Aharonov-Bohm caging~\cite{jv}. Thus, the quantum imprisonment induced by quantum interference may inspire the experimentalists to test our results.

Before ending this discussion, one important issue we should highlight a pertinent question that can we manipulate the current using the internal parameters of the system by including minimal asymmetry in our system? The deterministic off-diagonal anisotropy which is used in the previous section to show the sustainability of the extended modes, may control the magnitude of the current. This perception comes from an obvious standpoint. As we have already witnessed that off-diagonal anisotropy can modify the accessibility of the underlying system to the injected excitation. With this preliminary knowledge, we have checked the variation of magnitude of persistent current against the off-diagonal mismatch factor $r$ (Fig.~\ref{persist}(c)). From the diagram, we observe that the magnitude remarkably enhances with $r$. Therefore, one can control the current in a subtle way using the internal parameters of the Hamiltonian.

--------------------


\section{Analogous extention to Photonics}
\label{phon}
	
\begin{figure}[ht]
		\includegraphics[clip,width=0.44\textwidth]{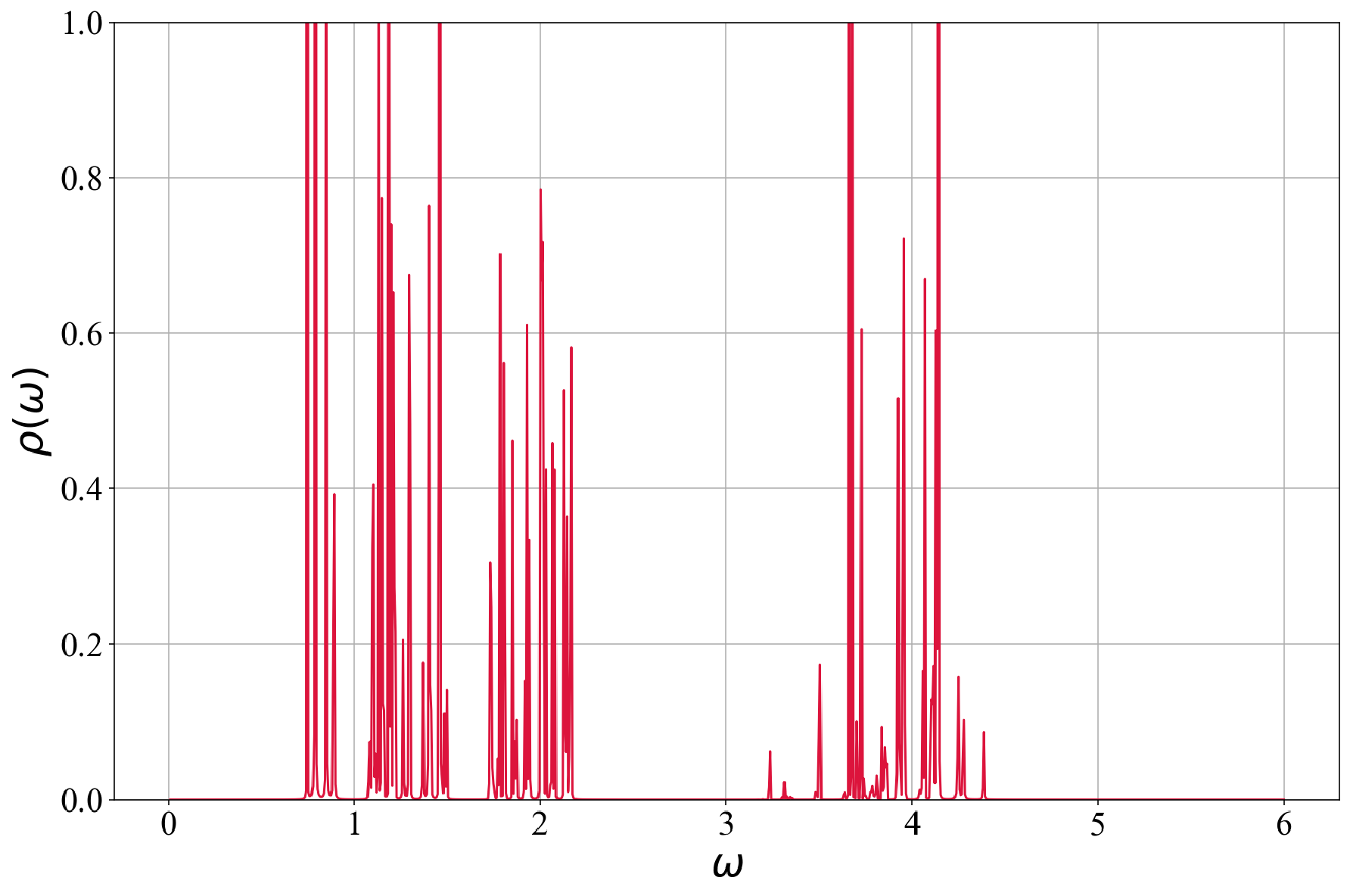}
		\caption{Variation of photonic modes with frequency for third generation of corral waveguide presenting the allowed modes.}
		\label{phn}
	\end{figure}

The	Proposition for geometry-induced localized state offered by this quantum corral can be extended for a photonic waveguide network arranged in the same form of the corral. The grafted photonic lattice can serve as the testing ground for the theoretical prediction presented so far. The wave propagation of classical wave in such single–mode waveguide may be described using the standard wave equation, viz.,
	
\begin{equation}
	\frac{\partial^2 \psi_{ij}}{\partial x^2}
	+ \frac{\omega^2 \epsilon_r}{c^2}\,\psi_{ij}(x) = 0
\end{equation}
The solution of the above equation takes the form,
	
\begin{equation}
\psi_{ij}(x) =
\psi_i \frac{\sin\!\big[k(l_{ij}-x)\big]}{\sin l_{ij}k}
+ \psi_j \frac{\sin(kx)}{\sin l_{ij}k}
\end{equation}
where, $k = \frac{\omega}{a}\sqrt{\epsilon_r}$. Flux conservation condition transform the
	above solution into a tight–binding analogue of the difference equation,
	\begin{equation}
		- \psi_i \sum_{j} \cot \theta_{ij}
		+ \sum_{j} \csc \theta_{j}\, \psi_j = 0
	\end{equation}

	where $\theta_{ij} = kl_{ij}$  being the length of the waveguide segment
	that can be set as uniform, \textit{a}, say. This mapping enables us to evaluate the
	frequency of incoming photon that will be eventually
	localized due to destructive quantum interference. For example,
	for a first generation corral waveguide the frequency corresponding
	to the bound state ($E = \epsilon - 2t$) is $f = \frac{c}{6a}\sqrt{\epsilon_r}$ ($\epsilon_r$ is diaelectric parameter). With the increase in hierarchy such non–diffusive frequencies will constitute the photonic density of states. The adjacent diagram (Fig.~\ref{phn}) depicts the variation of photonic density of modes with frequency for a third generation corral waveguide structure.
	 The detailed analysis will be published elsewhere. However, this concept, to our understanding, may make the photonic waveguide experimentalists curious.

\section{Summary}
A methodical spectral analysis for a \textit{self-similar} quantum corral is reported within the tight-binding framework. Absence of magnetic perturbation leads to controlled imprisonment of the electron of a definite energy which follows the same hierarchy as that of the geometry. Finite flux is able to bring phenomenal change in the response of the system. Flux induced phase coherence can invite \textit{absolutely continuous} band of extended states even for a non-translationally invariant system.
The flux can thus improve transparency of the system to the excitation by manipulating the band curvature (hence mobility) in a comprehensive style. The existence of diffusive electron states have been thoroughly verified with the evaluation of transport and inverse participation ratio. The states are found to be robust in respect of application of diagonal disorder or off-diagonal anisotropy. Finally, a detailed multifractal analysis indicates the appearance of single-particle mobility edge in the spectrum. The reflection of quantum interference has been manifested through the study of persistent current. Tunability of the current has been examined also. The entire spectral analysis may inspire the experimentalists to test the engineering of quantum states for this non-Euclidean fractal substrate.

	\label{closing}
	
	
		
		
	
	\begin{acknowledgments}
	Both SB and RA gratefully acknowledge the computational facility provided by the Department of Physics, Acharya Prafulla Chandra College.
	AN is grateful to Prof. Arunava Chakrabarti for his helpful suggestions regarding the results.

	\end{acknowledgments} 


\begin{thebibliography}{60}
	
	\bibitem{ander} P. W. Anderson, Phys. Rev. \textbf{109}, 1492 (1958).
	\bibitem{ander2} E. Abrahams, P. W. Anderson, D. C. Licciardello, and T. V. Ramakrishnan, Phys. Rev. Lett.
	\textbf{42}, 673 (1979).
	\bibitem{bor} R. E. Borland, Proc. R. Soc. Lond. A \textbf{274}, 274529 (1963).
	\bibitem{dey} F. Deylon, Y. Le\'{e}vy, and B. Souillard, J. Stat. Phys. \textbf{41}, 375 (1985).
	\bibitem{billy} J. Billy et. al. Nat. \textbf{453}, 895 (2008).
	\bibitem{kramer} B. Kramer and A. MacKinnon, Rep. Prog. Phys. \textbf{56}, 1469 (1993).
	\bibitem{van} A. Lagendijk, B. van Tiggelen, and D. S. Wiersma, Phys. Today \textbf{62}, 24, (2008).
	\bibitem{mp} M. P. Van Albada and A. Lagendijk, Phys. Rev. Lett.
	\textbf{55}, 2692 (1985).
	\bibitem{ds} D. S. Wiersma, P. Bartolini, A. Lagendijk, and R. Righini, Nature \textbf{390},
	671 (1997).
	\bibitem{martin} L. Martin, G. Di Giuseppe, A. Perez-Leija, R. Keil,
	F. Dreisow, M. Heinrich, S. Nolte, A. Szameit, A. F.
	Abouraddy, D. N. Christodoulides, and B. E. A. Saleh,
	Opt. Exp. \textbf{19}, 13636 (2011).
	\bibitem{segev} M. Segev, Y. Silberberg, and D. N. Christodoulides, Nat. Photonics \textbf{7}, 197 (2013).
	\bibitem{roati} G. Roati, C. D’Errico1, L. Fallani, M. Fattori, C. Fort1, M.
	Zaccanti1, G. Modugno1, M. Modugno, and M. Inguscio, Nature 453, 895 (2008).
	\bibitem{white} D. H. White, T. A. Haase, D. J. Brown, M. D. Hoogerland, M. S. Najafabadi, J. L. Helm, C. Gies, D. Schumayer, and D. A. W. Hutchinson, 
	Nat. Comm. \textbf{11}, 4942 (2020).
	\bibitem{schri} M. Schreiber and H. Grussbach, Phys. Rev. Lett. \textbf{76}, 1687 (1996).
	\bibitem{marko} I. Trav\'{e}nec and P. Marko\'{s}, Phys. Rev. B \textbf{65}, 113109 (2002).
	\bibitem{sacha} A. Kosior and K. Sacha, Phys. Rev. B \textbf{95}, 104206 (2017).
	\bibitem{cook} M. Brzezi\'{n}ska, A. M. Cook, and T. Neupert, Phys. Rev.
	B \textbf{98}, 205116 (2018).
	\bibitem{prem} S. Pai and A. Prem, Phys. Rev. B \textbf{100}, 155135 (2019).
	\bibitem{bp1} B. Pal and K. Saha, Phys. Rev. B \textbf{97}, 195101 (2018).
	\bibitem{an} A. Nandy, Phys. Scr. \textbf{96}, 045802 (2021).
	\bibitem{hana} H. Hanafi, P. Menz, and C. Denz, Adv. Opt. Mat.
	\textbf{10}, 2102523 (2022).
	\bibitem{xie} Y. Xie, L. Song, W. Yan, S. Xia, L. Tang, D. Song, J.-W.
	Rhim, and Z. Chen, APL Photonics \textbf{6}, 116104 (2021).
	\bibitem{song} L. Song, Y. Xie, S. Xia, L. Tang, D. Song, J.-W. Rhim,
	and Z. Chen, Laser Photonics Rev. \textbf{17}, 2200315 (2023).
	\bibitem{sb} S. Biswas and A. Chakrabarti, Physica E \textbf{153}, 115762
	(2023).
	\bibitem{hoo} M. Fremling, M. van Hooft, C. M. Smith, and L. Fritz,
	Phys. Rev. Research \textbf{2}, 013044 (2020).
	\bibitem{yuan} A. A. Iliasov, M. I. Katsnelson, and S. Yuan, Phys. Rev.
	B \textbf{101}, 045413 (2020).
	\bibitem{lage} L. L. Lage and A. Latg\'{e}, Phys. Chem. Chem. Phys. \textbf{24}, 19576 (2022).
	\bibitem{shang} J. shang et. al., Nat. Chem. \textbf{7}, 389 (2015).
	\bibitem{slot} S. N. Kempkes, M. R. Slot, S. E. Freeney, S. J. M. Zevenhuizen, D. Vanmaekelbergh, I. Swart, and C. M. Smith, Nat. Phys. \textbf{15}, 127 (2019).
	\bibitem{lust} Z. Yang, E. Lustig, Y. Lumer, and M. Segev, Light Sci.
	Appl. 9, 128 (2020).
	\bibitem{yori} H. Yorikawa, J. Phys. Comm. \textbf{3}, 085004 (2019).
	\bibitem{bp2} B. Pal, J. Appl. Phys. \textbf{137}, 144301 (2025).
	\bibitem{bp3} B. Pal, Phys. Stat. Solidi- Rapid Research Lett. \textbf{19}, 2500129 (2025).
	\bibitem{amun} M. Amundsen, V. Jurici\'{c}, and J. A. Ouassou, Appl. Phys. Lett. \textbf{125}, 092601 (2024).
	\bibitem{koch} G. Koch and A. Posazhennikova, Phys. Rev. A \textbf{110}, 033301 (2024).
	\bibitem{science} G. R. Newkome e. al., Science \textbf{312}, 1782 (2006).
	\bibitem{carbon} L. L. Lage and A. Latg\'{e}, Front. Carbon \textbf{2}, 1305515 (2023).
	\bibitem{acsimplex} A. Chakrabarti, Phys. Rev. B \textbf{72}, 134207 (2005).
	\bibitem{acnew} S. Biswas and A. Chakrabarti, Phys. Rev. B \textbf{108}, 125430 (2023).
	\bibitem{skrk} S. Kirkpatrick and T. P. Eggarter, Phys. Rev. B 6, 3598
	(1972)
\bibitem{Mujica1994}V.~Mujica, M.~Kemp, and M.~A.~Ratner, J.~Chem.~Phys. \textbf{101}, 6849 (1994).
\bibitem{Dutta2010} P.~Dutta, S.~K.~Maiti, and S.~N.~Karmakar, Org.~Elect. \textbf{11}, 1120 (2010).
\bibitem{Dutta2013} P.~Dutta, S.~K.~Maiti, and S.~N.~Karmakar, J.~Appl.~Phys. \textbf{114}, 034306 (2013).

\bibitem{3}Phys. Rev. Lett. 81, 5888 (1998), Phys. Rev. Lett.\textbf{85}. 
\bibitem{citekey}3906 (2000), Phys. Rev. B 64.
\bibitem{}3155306 (2001), Phys. Rev. Lett. 88, 227005 (2002)).
\bibitem{fe} F. Evers and A. D. Mirlin, Rev. Mod. Phys. \textbf{80}, 1355 (2008).
\bibitem{hh} H. Hiramoto and M. Kohmoto, Phys. Rev. Lett. \textbf{62}, 2714
(1989).
\bibitem{lj} L. J. Vasquez, A. Rodriguez, and R. A. Römer, Phys. Rev. B \textbf{78},
195106 (2008).
\bibitem{cg} C. Godreche and J. M. Luck, J. Phys. A: Gen. Phys.\textbf{23}, 3769
(1990).
\bibitem{hih} H. Hiramoto and M. Kohmoto, Phys. Rev. B \textbf{40}, 8225 (1989).
\bibitem{gef} H.-F. Cheung, Y. Gefen, E. K. Riedel, and W.-H. Shih Phys. Rev. B \textbf{37}, 6050 (1988).
\bibitem{jv} J.Vidal, R.Mosseri, and B. Doucot Phys. Rev. Lett.\textbf{81}, 5888 (1998)



	
\end{thebibliography}
	\end{document}